\documentclass[aps, showpacs, preprintnumbers,  nofootinbibt,twocolumn,referee]{revtex4}
\usepackage{amsmath}
\usepackage{eurosym}
\usepackage{amssymb}
\usepackage{graphicx}
\usepackage{color}

\def\be{\begin{equation}}
\def\ee{\end{equation}}
\def\bea{\begin{eqnarray}}
\def\eea{\end{eqnarray}}

\begin{document}

\title{Slowly rotating Bose Einstein Condensate galactic dark matter halos, and their rotation curves}

\author{Xiaoyue Zhang}
\email{zhangxiaoyue23@pku.edu.cn}
\affiliation{School of Physics and Yat Sen School, Sun Yat-sen University, \\
Guangzhou 510275, People's Republic of China}
\affiliation{Department of Astronomy, Peking University, Beijing 100871, People's Republic of China}
\author{Man Ho Chan}
\email{chanmh@eduhk.hk}
\affiliation{Department of Science and Environmental Studies, The Education University of Hong Kong, \\
Hong Kong, People's Republic of China}
\author{Tiberiu Harko}
\email{t.harko@ucl.ac.uk}
\affiliation{Department of Physics, Babes-Bolyai University, Kogalniceanu Street,
Cluj-Napoca 400084, Romania,}
\affiliation{School of Physics, Sun Yat-sen University,
Guangzhou 510275, People's Republic of China}
\affiliation{Department of Mathematics, University College London, Gower Street, London
WC1E 6BT, United Kingdom}
\author{Shi-Dong Liang}
\email{stslsd@mail.sysu.edu.cn}
\affiliation{State Key Laboratory of Optoelectronic Material and Technology,
and Guangdong Province Key Laboratory of Display Material and Technology, \\
Sun Yat-Sen University, Guangzhou 510275, People's Republic of China}
\author{Chun Sing Leung}
\email{chun-sing-hkpu.leung@polyu.edu.hk}
\affiliation{Department of Applied Mathematics, Hong Kong Polytechnic University, \\
Kowloon, Hong Kong, People's Republic of China}

\date{\today }

\begin{abstract}
If dark matter is composed of massive bosons, a Bose-Einstein Condensation process must have occurred during the cosmological evolution. Therefore galactic dark matter may be in a form of a condensate, characterized by a strong self-interaction. We consider the effects of rotation on the Bose-Einstein Condensate dark matter halos, and we investigate how rotation might influence their astrophysical properties. In order to describe the condensate we use the Gross-Pitaevskii equation, and the Thomas-Fermi approximation, which predicts a polytropic equation of state with polytropic index $n=1$. By assuming a rigid body rotation for the halo, with the use of the hydrodynamic representation of the Gross-Pitaevskii equation we obtain the basic equation describing the density distribution of the rotating condensate. We obtain the general solutions for the condensed dark matter density, and  we derive the general representations for the
mass distribution, boundary (radius), potential energy, velocity dispersion, tangential velocity and for the logarithmic density and velocity slopes, respectively. Explicit expressions for the radius, mass, and tangential velocity are obtained in the first order of approximation, under the assumption of slow rotation. In order to compare our results with the observations we fit the theoretical expressions of the tangential velocity of massive test particles moving in rotating Bose-Einstein Condensate dark halos with the data of 12 dwarf galaxies, and the Milky Way, respectively.
\end{abstract}


\maketitle

\tableofcontents

\section{Introduction}

The assumption of the existence of dark matter (DM) is one of the cornerstones of present day cosmology and astrophysics \cite{d1,d2,d3,d4,d5} . The first evidence for its presence in the Universe was provided by the study of the galactic rotation curves. More exactly, the idea of DM was first suggested to explain the rotation curves of spiral galaxies, whose rotation curves decay far more slowly than one would expect by taking into account the effects of baryonic matter (gas and stars) only. This behavior is considered as evidence for the existence of a supplementary (and exotic) missing mass
component,  most likely consisting of new particle(s) lying beyond the standard model of particle physics. The rotation curves still represent the most powerful and convincing evidence for DM \cite{Persic,Read,Salucci,Haghi}.  But various other astrophysical and cosmological observations have also provided evidence for the existence of dark matter, like, for example, the recent determination of the cosmological parameters from the Planck satellite observations of the Cosmic Background Radiation \cite{Planck}.  These observations have also shown that dark matter cannot be explained by  baryonic matter only, thus confirming the standard $\Lambda $ Cold Dark Matter ($\Lambda$CDM) cosmological paradigm. Other types of observations, such as gravitational lensing also require the existence of dark matter for a consistent interpretation of the data \cite{Wegg, Munoz, Chuda}. A particularly strong evidence for the existence of dark matter is provided by the observations of a galaxy cluster called the Bullet Cluster. In this cluster  the baryonic matter and the dark matter components are separated due of a collision of its two components that occurred in the past \cite{massey2007dark}. Measurements of the cosmological parameters by using the Planck data on the cosmic microwave background radiation indicate that the Universe is composed of 4\% baryons, 22\% non-baryonic dark matter and 74\% dark energy\cite{Planck}.

Dark matter models can be divided into three types, cold, warm and hot dark matter models, respectively, by the energy of the particles composing them \cite{Overduin}.  The main candidates for dark matter are the WIMPs (Weakly Interacting Massive Particles) and the axions \cite{Overduin}. WIMPs are heavy particles that interact via the weak force \cite{Cui,Matsumoto}. Axions are bosons that were first proposed to solve the strong CP problem \cite{Mielke, Schwabe}. If the axions form the dark matter, then at low temperature the axion gas will form a Bose-Einstein Condensate (BEC).

There are also other theories that try to explain the observations without introducing dark matter. These theories are based on a modification of the law of gravity at the galactic scales. The earliest one of them is the MOND theory (Modified Newtonian Dynamics) \cite{Milgrom}. Modified gravity theories have also been used extensively as an alternative to dark matter \cite{alt1,alt2,alt3,alt4,alt5,alt6,alt7,alt8,alt9,alt10,alt11}. An interesting possibility to detect the presence of dark matter is via its possible annihilation.  If such a physical process does indeed occur, then a large numbers gamma ray photons  and positrons could be produced. Observationally, some excess positron emission in our galaxy has been detected \cite{Chann1,Chann2, Chann3,Chan1,Chan2,Chan3,Chan4,Chan5}. Therefore, it may be possible that the excess positron and gamma-ray emissions could be explained by the annihilation of dark matter
with mass $m \sim 10 - 100$  GeV \cite{Chann1,Chann2,Chann3}. For a detailed discussion of this problem, as well as of the possibility of alternative interpretations of the observational data see \cite{Chan1,Chan2,Chan3,Chan4,Chan5}.

Even that dark matter models can give a good explanation of the qualitative behavior and constancy of the rotation curves, an important contradiction did arise as a result of the in depth comparison of the simulation results with the observations. Data on almost all observed rotation curves show that they rise less steeply than cosmological simulations of structure
formation in the standard $\Lambda$CDM model in presence of a single pressureless dark matter component do predict.
The simulations indicate a central dark matter density profile that behaves as $\rho \sim 1/r$ (a cusp) \cite{NFW}, while the observed rotation curves indicate the presence of constant density cores \cite{H,OH}.  This is the so-called core-cusp problem in dark matter physics. Another important open question dark matter models have to face is the "too big to fail" problem \cite{Boylan1,Boylan2}. By using the Aquarius simulations it was shown that the most massive subhalos in the dark matter halos predicted in the $\Lambda$CDM model are inconsistent with the dynamics of the brightest Milky Way dwarf spheroidal galaxies \cite{Boylan2}. While the best-fitting hosts of the dwarf spheroidals galaxies have $12 < V_{max} < 25$ km/s, the $\lambda$CDM simulations all predict at least ten subhalos with $V_{max} > 25$ km/s. These results cannot be explained in the framework of the  $\Lambda$CDM-based models of the satellite population of the Milky Way. The main problem emerging here is related  to the densities of the satellites, with the dwarf spheroidals required to have dark matter halos that are a factor of $\sim$5 more massive than it is observed.

These problems related to the physics of the dark matter may be solvable if one goes one step beyond the standard $\Lambda $CDM model, and assume that the dark matter particles may posses some forms of self-interaction. Such a possibility has gained some observational support after the study of the data provided by the  observations of 72 cluster collisions, including both `major' and `minor' mergers, with the observations done by  using the Chandra and Hubble Space Telescopes \cite{Harvey}. Collisions between galaxy clusters can provide an important test of the non-gravitational forces acting on dark matter, and the analysis done in \cite{Harvey} gives an upper limit of the ratio of the self-interaction cross-section $\sigma _{DM}$ and of the  mass $m$ of the dark matter particle as $\sigma_{DM}/m < 0.47$ cm$^2$/g (95\% Confidence Level). A new upper limit on the self-interaction cross-section of dark matter of $\sigma_{DM}<1.28$ cm$^2$/g (68\% Confidence Level), was obtained in \cite{Jauzac}. From a theoretical point of view different self-interacting
dark matter models were investigated in \cite{Carlson,Laix,Saxton1,Saxton2}. The effects of self-interacting dark matter on the tidal stripping and evaporation of satellite galaxies in a Milky Way-like host were investigated in \cite{Dooley}. Velocity-independent self-interacting dark matter models show a modest increase in the stellar stripping effect with satellite mass, whereas velocity-dependent self-interacting dark matter models show a large increase in this effect towards lower masses, making observations of ultra-faint dwarfs prime targets for distinguishing between and constraining self-interacting dark matter models. The response of self-interacting dark matter  halos to the growth of galaxy potentials using idealized simulations, each run in tandem with standard collisionless Cold Dark Matter (CDM) was investigated in \cite{Elbert}. A greater diversity in the self-interacting dark matter halo profiles was found, as compared to the standard CDM halo profiles. A self-interacting dark matter halo simulated with cross section over mass $\sigma _{DM}/m=0.1$ cm$^2$/g provides a good match to the measured dark matter density profile of A2667, while an adiabatically-contracted CDM halo is denser and cuspier. The cored profile of the same halo simulated with $\sigma _{DM}/m=0.5$ cm$^2$/g is not dense enough to match A2667.  These findings are in agreement with previous results \cite{Harvey} that $\sigma _{DM}/m\geq 0.1$ cm$^2$/g is disfavored for dark matter collision velocities in excess of about 1500 km/s. Therefore the possibility that dark matter is a self-interacting component of the Universe cannot be rejected {\it a priori}, and physical models whose component particles  naturally exhibit this property may provide valuable explanations and suggestions for the dark  matter candidates, and for their properties. From both a fundamental theoretical point of view, as well as  from a phenomenological
perspective, {\it the physically best motivated} self-interacting dark matter model can be obtained by assuming that dark matter is in a
{\it Bose-Einstein Condensate phase}.

The idea that at very low temperatures all integer spin particles may occupy the lowest quantum state, at which point macroscopic quantum phenomena become apparent, was proposed,  from a statistical
physical point of view by Bose and Einstein in the 1920s \cite{Bose,Ein,Ein1}.  The Bose-Einstein Condensation process is determined by the quantum mechanical  correlation of the gas particles, which implies that the de Broglie thermal wavelength is
greater than the mean interparticle distance. The transition to the condensate phase begins when the temperature $T$ of the boson gas is lower than the critical one, $T_{cr}$, given by \cite{Dalfovo,Pit,Pethick,Gr}
\begin{equation}  \label{Ttr}
T_{cr}=\frac{2\pi\hbar^2\rho_{cr}^{2/3}}{ \zeta^{2/3}(3/2)m^{5/3}k_B},
\end{equation}
where $m$ is the particle mass in the condensate, $\rho_{cr}$ is the
critical transition density, $k_B$ is Boltzmann's constant, and $\zeta $ denotes the
Riemmann zeta function.

It took around seventy years to achieve the experimental
realization of the Bose-Einstein Condensation, which was first observed in dilute alkali gases in 1995 \cite{exp1,exp2,exp3}.  From a physical point of view the presence of a  BEC in an experimental framework is indicated  by the appearance of sharp peaks in both coordinate and momentum space
distributions of the particles.

Up to now, the main evidence for the existence of BECs comes from laboratory experiments, performed on a very
small scale. However, the possibility of the presence of some forms of
condensates in the cosmic environment cannot be rejected a priori, and the implications of the possible existence of a condensate state of matter in a astrophysical or cosmological background is certainly worth to investigate. It has been hypothesized that due to their superfluid properties
in general relativistic compact objects, like neutron or quark stars, the neutrons may form  Cooper pairs, which would condense eventually. Bose-Einstein Condensate stars could have maximum masses of the order of 2 $M_{\odot}$,
maximum central densities of the order of $0.1-0.3\times 10^{16}$ g/cm$^3$,
and minimum radii in the range of 10-20 km, respectively. Their interesting physical and astrophysical properties were investigated in
\cite{starsm1, stars0, stars1,stars2,stars3,stars4,stars5,stars6,stars7, stars8}.

The idea that dark matter is in the form of a Bose-Einstein Condensate was
proposed initially in \cite{early1}, and then rediscovered/reinvestigated, in \cite{early2, early3,early4,early5, early6,early7, early8, early9a, silverman2002dark, rotha2002vortices, early9}.
A systematic study of the properties of the BEC dark matter halos, based on the non-relativistic Gross-Pitaevskii (GP) equation in the presence of a confining gravitational potential,  was initiated in
\cite{BoHa07}.  A further simplification of the
mathematical formalism of the gravitationally bounded BECs can be achieved by introducing the Madelung
representation of the wave function, which allows the representation of the GP
equation in the equivalent form of a continuity equation, and of a hydrodynamic Euler type
equation. Hence with the use of the Madelung representation we obtain the fundamental result that dark matter
can be described as a non-relativistic, Newtonian Bose-Einstein Condensed gas in a gravitational trapping potential, with the pressure and density obeying a polytropic equation of state, with polytropic index $n=1$. The validity of the BEC dark
matter model was tested by fitting the Newtonian tangential velocity
equation to a sample of rotation curves of low surface
brightness and dwarf galaxies, respectively.

The thermal correction to the dark matter density profile where obtained in \cite{HaM}. In \cite{Har1} it was shown that the
density profiles of the Bose-Einstein Condensed dark matter generally show
the presence of an extended core, whose formation is explained by the strong
interaction between dark matter particles. A further observational test of the model can be obtained by computing the mean value of the logarithmic
inner slope of the mass density profile of dwarf galaxies, and by comparing it with the observations. The study of the properties of the Bose - Einstein
Condensate dark matter on a cosmological and astrophysical scales is presently a very active field of research \cite{inv0, inv1, inv2,inv3,inv4,inv5,inv6,inv7,inv8,inv9,inv10,inv11,inv12,inv13,inv14,inv15,inv16,inv17,inv18,inv19,inv20,inv21,inv22,inv23,inv24,inv25,inv26, inv27,inv28,inv29,inv30,inv31,inv32,inv33,inv34,inv35, inv36,inv37}. The properties of the Fuzzy Dark Matter, assumed to be formed of a light ($m\sim 10^{-22}$ eV) boson having a de Broglie wavelength $\lambda \sim 1$ kpc, were recently investigated in \cite{Hui}.

If the static properties of the BEC dark matter halos have been studied extensively, their rotational properties have attracted less attention. In \cite{early9a} the presence of vortices in a self gravitating BEC dark halo, consisting of ultra-low mass scalar bosons was investigated, and it was pointed out that rotation of the dark matter imprints a background phase gradient on the condensate, which induces a harmonic trap potential for vortices.  A detailed study of the vortices in rotating BEC dark matter halos was performed in \cite{inv3}, where strong bounds for the boson mass, interaction strength, the shape and quantity of vortices in the halo, and the critical rotational velocity for the nucleation of vortices were found. An exact solution for the mass density of a single, axisymmetric vortex was also found. The effects of rotation on a superfluid BEC dark matter halo were explored in \cite{inv4}, by assuming that a vortex lattice forms. With fine-tuning of the bosonic particle mass and the two-body repulsive interaction strength, it was  found that sub-structures on rotation curves that resembles some observations in spiral galaxies could exist. The study of the equilibrium of self-gravitating, rotating BEC haloes, which satisfy the Gross-Pitaevskii-Poisson equations was performed in \cite{inv15}. Vortices are expected to form for a wide range of BEC
parameters. However, vortices cannot form for vanishing self-interaction.  The question if and when vortices are energetically favored was also considered, and it was found that  vortices form as long as self-interaction is strong enough.

 It is the goal of the present paper to study the properties of the BEC dark matter halos in the presence of rotation. Rotation might be a general property of galaxies, whose origin may be traced back to some physical processes in the early Universe. In order to describe the Bose Einstein Condensate dark matter we adopt the Gross-Pitaevskii equation, which gives an effective  mean-field description of the multi-particle bosonic system. The mathematical description of the condensate is significantly simplified after introducing the Thomas-Fermi approximation, which allows the description of the dark matter as a gas obeying a polytropic equation of state, with polytropic index $n=1$. By assuming a rigid body rotation for the halo, with the use of the hydrodynamic representation of the Gross-Pitaevskii equation, we obtain the basic relation describing the density distribution of the rotating condensate, which naturally generalizes the previously obtained static profile. From the density distribution of the rotating condensate we derive the general representations for the
mass distribution, boundary (radius), potential energy, velocity dispersion, tangential velocity,  as well as for the logarithmic density and velocity slopes. From the general results we obtain explicit expressions for the radius, mass, and tangential velocity in the first order of approximation, under the assumption of slow rotation. A comparison of our results with the observations is performed by fitting the theoretical expressions of the tangential velocity of massive test particles moving in rotating Bose-Einstein Condensate dark halos with the data of 12 dwarf galaxies, and of the Milky Way galaxy, respectively.

The present paper is organized as follows. The mathematical and physical description of the Bose-Einstein Condensate dark matter is introduced in Section~\ref{sect1}, where the Gross-Pitaevskii equation, and the Thomas-Fermi approximation are presented. The rotating Bose-Einstein Condensate dark matter structures are investigated in Section~\ref{sect2}, by using the general approach that allows us to obtain the exact expression of density as expressed in terms of Legendre polynomials. The slowly rotating dark matter halo in the framework of the  first order approximation is also considered, and the density profile, as well as the radius are also presented. The gravitational and astrophysical properties of the rotating Bose-Einstein Condensate dark matter halos are studied in Section~\ref{sect3}. In this Section we derive the expressions of a number of important astrophysical quantities, like, for example, the mass distribution, potential energy, logarithmic slopes of the density and velocity, which could allow an in depth comparison between the theoretical model and observations. The fitting of the theoretical model with astronomical/astrophysical data is performed in Section~\ref{sect4},  where we compare the predicted Bose-Einstein Condensate galactic rotation curves with the observational data from 12 dwarf galaxies, and the Milky Way galaxy. We discuss and conclude our results in Section~\ref{sect5}.

\section{The Bose-Einstein Condensate dark matter model}\label{sect1}

In the present Section we briefly introduce the fundamental physical concepts related to the Bose-Einstein Condensation, as well as the basic equations describing the rotating condensate. It has been shown that if the dark matter is composed of ultralight boson particles with mass $m\sim10^{-22}\;{\rm eV}$ and wavelength $\lambda \sim 1$ kpc, then the transition temperature to a Bose-Einstein Condensate is of the order $T_c\sim10^9\;{\rm K}$  \cite{silverman2002dark}.  Hence, if dark matter is composed of Bose particles, like, for example, the axion, it is quite natural to assume that dark matter is in a Bose-Einstein Condensate state. For a recent discussion of the arguments from particle physics that may motivate the existence of the  ultra-light dark matter, as well as of its properties and astrophysical signatures see \cite{Hui}.

\subsection{The Gross-Pitaevskii Equation}

A Bose-Einstein Condensate is a phase of matter in which all the particles are localized in the ground state. The Bose-Einstein Condensation  occurs for particles that have integer spins, and obey the Bose-Einstein statistics.
We will consider in the following that the bosons are weakly interacting, and that their interaction is described by a two-body interparticle potential. We start our analysis by writing down first the many-body Hamiltonian of the bosonic system,
\begin{equation}
\begin{aligned}
\hat{H}
=&\int\mathrm{d}\boldsymbol{r}\hat{\Psi}^\dag(\boldsymbol{r})\Big[-\frac{\hbar^2}{2m}\nabla^2+V_{rot}(\boldsymbol{r})
  +V_{ext}(\boldsymbol{r})\Big]\hat{\Psi}(\boldsymbol{r})\\
&+\frac 12\int\mathrm{d}\boldsymbol{r}\mathrm{d}\boldsymbol{r'}\hat{\Psi}^\dag(\boldsymbol{r})\hat{\Psi}^\dag(\boldsymbol{r'})
  V(\boldsymbol{r}-\boldsymbol{r'})\hat{\Psi}(\boldsymbol{r})\hat{\Psi}(\boldsymbol{r'}),
\end{aligned}
\end{equation}
where $\hat{\Psi}(\boldsymbol{r})$ and $\hat{\Psi}^\dag(\boldsymbol{r'})$ are the annihilation operator and the creation operator at the position $\boldsymbol{r}$, respectively, $V$ denotes the interparticle interaction potential, and $m$ is the mass of the particle in the condensate. In the case of a rotating dark matter halo, the external potential $V_{ext}$ is the gravitational potential, and $V_{rot}$ is the effective centrifugal potential. In the following we will adopt the comoving frame, that is, the frame that's rotating with the same speed as the system.

To simplify the mathematical formalism, we introduce the mean field description, in which we decompose the field operator in the form $\hat{\Psi}(\boldsymbol{r})=\Psi_0+\hat{\Psi}'(\boldsymbol{r})$, and treat the operator $\hat{\Psi}'(\boldsymbol{r})$ as a small perturbation. Then for the mean field component $\Psi_0$ we have  $\Psi_0=\sqrt{N/V}$, where $N$ is the total particle number, and $V$ is the volume. Hence $\Psi_0$ is equal to the square root of the number density of the particles \cite{Dalfovo}.

In the general time-dependent case, the field operator in the Heisenberg picture is given by
\begin{equation}
  \hat{\Psi}(\boldsymbol{r},t)=\psi(\boldsymbol{r},t)+\hat{\Psi}'(\boldsymbol{r},t),
\end{equation}
where $\psi(\boldsymbol{r},t)=\langle\hat{\Psi}(\boldsymbol{r},t)\rangle$ is also called the condensate wave function. Then for the number density of the condensate we have $\rho_n(\boldsymbol{r},t)=|\psi(\boldsymbol{r},t)|^2$. The normalisation condition is $N=\int\rho_n(\boldsymbol{r},t)d^3\boldsymbol{r}$.

In the Heisenberg representation the equation of motion of the field operator is
\begin{equation}\label{gpeq}
\begin{aligned}
  i\hbar\frac{\partial}{\partial t}\hat{\Psi}(\boldsymbol{r},t)=&[\hat{\Psi},\hat{H}]
  =\Big[-\frac{\hbar^2}{2m}\nabla^2+V_{rot}(\boldsymbol{r})
  +V_{ext}(\boldsymbol{r})\\
 &+ \int\mathrm{d}\boldsymbol{r'}\hat{\Psi}^\dag(\boldsymbol{r'},t)
  V(\boldsymbol{r}-\boldsymbol{r'})\hat{\Psi}(\boldsymbol{r'},t)\Big]\hat{\Psi}(\boldsymbol{r},t).
\end{aligned}
\end{equation}

In the theory of the Bose-Einstein Condensation one usually assumes that the interparticle interaction is a short range interaction, and hence we can write the interaction potential as being proportional to a constant, which is related to the scattering length, times a Dirac delta function \cite{barcelo2001analogue}:
\begin{equation}
\begin{aligned}
V(\boldsymbol{r}'-\boldsymbol{r})=&\lambda\delta(\boldsymbol{r}'-\boldsymbol{r}),
\end{aligned}
\end{equation}
with
\be
\lambda=\frac{4\pi a\hbar^2}m,
\ee
where $a$ is the scattering length. To obtain a more general description, in the following we introduce the function $g(|\psi(\boldsymbol{r},t)|^2)$ to describe the self-interaction term \cite{BoHa07}. In the standard approach to Bose-Einstein Condensation the self-interaction is assumed to have a quadratic form,  so that $g(|\psi(\boldsymbol{r},t)|^2)=\frac12\lambda|\psi(\boldsymbol{r},t)|^4$ \cite{Dalfovo}.

With this approximation of the potential, and with the use of the mean field approximation, by integrating over Eq.(\ref{gpeq}) we obtain the Gross-Pitaevskii equation, describing the main properties of a Bose-Einstein Condensate, as
\bea
  i\hbar\frac{\partial}{\partial t}\psi(\boldsymbol{r},t)&=&\Big[-\frac{\hbar^2}{2m}\nabla^2+V_{rot}(\boldsymbol{r})
  +V_{ext}(\boldsymbol{r})\nonumber\\
 &&+ g'(|\psi(\boldsymbol{r},t)|^2)\Big]\psi(\boldsymbol{r},t).
\eea

To give a more direct physical interpretation of the Gross-Pitaevskii equation, we introduce Madelung representation of the wave function,
\begin{equation}
  \psi(\boldsymbol{r},t)=\sqrt{\rho_n(\boldsymbol{r},t)}\ e^{\frac i{\hbar}S(\boldsymbol{r},t)}
\end{equation}
which separates the wave function into two components, its magnitude, and a phase factor, respectively. The function $S(\boldsymbol{r},t)$ has the dimensions of an action. Then in the Madelung representation we have \cite{Gr}
\bea
&&\frac{i\hbar}{\psi}\frac{\partial\psi}{\partial t}=-\frac{\partial S}{\partial t}+\frac{i\hbar}{2\rho_n}\frac{\partial\rho_n}{\partial t}
 \frac1{\psi}\Bigg[-\frac{\hbar^2}{2m}\nabla^2+V_{rot}  +V_{ext}+ \nonumber\\
 && g'(|\psi|^2)\Bigg]\psi=
  -\frac{\hbar^2}{2m}\frac{\nabla^2\sqrt{\rho_n}}{\sqrt{\rho_n}}+\nonumber
 \frac1{2m}|\nabla S|^2+V_{rot}+V_{ext}\nonumber\\
 &&+g'(|\psi|^2)-\frac{i\hbar}{2\rho_n}(\nabla\rho_n\cdot\nabla S+\rho_n\nabla^2 S).
\eea

Then the Gross-Pitaevskii equation is separated into two parts. From the imaginary part we obtain,
\begin{equation}
  \frac{\partial\rho_n}{\partial t}+\nabla\cdot(\rho_n\boldsymbol{v})=0,
\end{equation}
where $\boldsymbol{v}=\frac{\nabla S}m$ is the velocity of the quantum fluid. This is the continuity equation. On the other hand, from the real part we obtain the equation
\bea\label{Euler}
&&  \rho_n\Big(\frac{\partial\boldsymbol{v}}{\partial t}+\boldsymbol{v}\times(\nabla\times\boldsymbol{v})+(\boldsymbol{v}\cdot\nabla)\boldsymbol{v}\Big)=-\nabla P(\rho_n)\nonumber\\
&&-\rho_n\nabla V_{rot}  -\rho_n\nabla V_{ext}-\nabla\cdot\sigma^Q,
\eea
which is the momentum conservation, or the hydrodynamic Euler equation \cite{inv29}. In Eq.~(\ref{Euler}) $P$ is the thermodynamic pressure, which is related to the mass density $\rho=\rho_nm$ of the condensate by a barotropic type equation of state  \cite{barcelo2001analogue, BoHa07}
 \be
 P(\rho)=g'\left(\frac{\rho}{m}\right)\frac{\rho}{m}-g\left(\frac{\rho}{m}\right).
 \ee

 The term $\sigma^Q$ is given by $\sigma ^{Q}=-\frac{\hbar^2}{2m}\frac{\nabla^2\sqrt{\rho}}{\sqrt{\rho}}$, and its divergence is called the quantum stress tensor.

Therefore the two equations describing the evolution of a Bose-Einstein Condensate are the continuity and the Eular equations of classical fluid dynamics, which describe viscosity free flows. Also, we can see from the definition of the velocity that the flow must be automatically irrotational. We will discuss this issue later.

\subsection{Thomas-Fermi Approximation}

When the number of particles in the condensate get large enough, the contribution to the energy of the quantum pressure term $\nabla\cdot\sigma^Q$ can be neglected except near the boundary \cite{Pethick}. Then the equations describing the condensate become purely classical in their mathematical form, even that their physical interpretation must be given in the framework of quantum statistical physics.

If we consider the Bose-Einstein Condensate to be static, or have a rigid body rotation, and we work in the corotating frame, then $\boldsymbol{v}=0$. Thus from Eq.~(\ref{Euler})  we obtain
\begin{equation}\label{13}
  \nabla^2\left[h(\rho )+V_{rot}+V_{ext}\right]=0,
\end{equation}
where $\nabla h(\rho )=\left(1/\rho\right)\nabla P(\rho)$. For the exterior potential we assume that it is the gravitational potential, $V_{ext}=V_{grav}=V$, and that it satisfies the Poisson equation,
\begin{equation}
\nabla^2V=4\pi G\rho,
\ee
where $G$ is the gravitational constant. For the rotational potential we adopt the expression
\be
V_{rot}=-\frac12\omega^2(x^2+y^2),
\end{equation}
where $\omega$ is the angular velocity of the dark matter halo.
In the case of the quadratic nonlinearity  we have $g'(\rho)=\lambda\rho/m$, and $g(\rho)=\lambda\rho^2/(2m^2)$, respectively. Thus for the equation of state of the condensate we obtain
\begin{equation}
P(\rho)=\frac{\lambda}{2m^2}\rho^2,
\ee
giving
\be
 \frac1{\rho}\nabla^2(P)=\frac{\lambda}{m^2}\nabla^2\rho.
\end{equation}
From the equation of state of the condensate we can see that $P\propto\rho^2$, and since for a general polytropic equation of state $\Gamma=1+1/n=2$, it follows that the polytropic index of the condensate is $n=1$.

Hence with the use of the equation of state of the BEC dark matter from Eq.~(\ref{13}) we obtain the equation describing the variation of the density of the rotating dark matter halo as
\begin{equation}\label{Helm}
  \nabla^2\rho+k^2\left(\rho-\frac{\omega^2}{2\pi G}\right)=0,
\end{equation}
where
\be
k=\sqrt{\frac{4\pi Gm^2}{\lambda}}=\sqrt{\frac{Gm^3}{a\hbar ^2}}.
\ee
 Eq.~(\ref{Helm}) is the Helmholtz equation. If the system had a different polytropic index $n\neq 1$, instead, we will get a general Lane-Emden equation \cite{BoHa07}, which will be nonlinear.

%For the Helmholtz equation, we can write down its general solution in the spherical coordinate,
%\begin{equation}
%  \rho(r,\theta)=\frac{\omega^2}{2\pi G}+\sum_{l=0}^{\infty}A_{2l}j_{2l}(kr)P_{2l}(\cos{\theta})
%\end{equation}

If the halo is nonrotating, $\omega =0$, and, under the assumption of spherical symmetry, Eq.~(\ref{Helm}) has the solution \cite{BoHa07}
\begin{equation}
\rho(r)=A_0\frac{\sin kr}{kr},
\end{equation}
where $A_0$ is an integration constant. One can obtain the radius $R$ of the static halo from the boundary condition $\rho(R)=0$, which gives
\bea
\hspace{-1cm}R&=&\frac{\pi}k=\pi \sqrt{\frac{a\hbar ^2}{Gm^3}}=\nonumber\\
\hspace{-0.5cm}&&13.5\times \left(\frac{a}{10^{-17}\;{\rm cm}}\right)^{1/2}\times \left(\frac{m}{10^{-36}\;{\rm g}}\right)^{-3/2}\;{\rm kpc}.
\eea
The central density is $\rho_c=\rho(0)=A_0$. We can see from the expression of the density that the radius only depends on the mass and scattering length of the particles. The size of the halo is independent of the central density.
We can determine $\rho_c$ from the normalisation condition, $\int d^3\boldsymbol{r}\rho=M$, thus obtaining
\be
\rho_c=A_0=\frac{Mk^3}{4\pi^2}=\frac{\pi M}{4R^3}=\frac{M}{4\pi^2}\left(\frac{Gm^3}{a\hbar ^2}\right)^{3/2}.
\ee

Once the BEC dark matter
density profile is known, all the global parameters of the BEC dark matter halo
(mass, radius, central density), as well as the rotational speeds of particles in stable circular motion can
be obtained in an exact form. These results open the possibility of the observational test of
the BEC dark matter model.

\subsection{Emergence of Vortices}

We have already mentioned that from the definition of the velocity $\boldsymbol{v}=\nabla S$, the flow must automatically satisfy the condition $\nabla\times\boldsymbol{v}=0$, and hence the quantum BEC fluid motion should be irrotational. However, if we expect the dark matter halo to be rotating, we must introduce singularities of the vorticity, and therefore the halo will contain a vortex lattice.

First, let us recall the concepts of vortex and vorticity. A vortex is a region of fluid in which the flow is rotating around an axis line. Its vorticity is the curl of the velocity, $\boldsymbol{w}=\nabla\times\boldsymbol{v}$. If the fluid rotates like a rigid body with an angular velocity $\boldsymbol{\Omega}=(0,0,\Omega)$, we have the velocity $\boldsymbol{v}=\boldsymbol{\Omega}\times\boldsymbol{r}=(-\Omega y,\Omega x,0)$ and the vorticity $\boldsymbol{w}=\nabla\times\boldsymbol{v}=2\boldsymbol{\Omega}$. A vortex can also be irrotational, if it has angular velocity $\boldsymbol{\Omega}=(0,0,\alpha r^{-2})$, its vorticity will be $0$ except at the axis line, where the vorticity will be infinite. If a fluid is irrotational, then although the particles have an angular velocity, they will not rotate over themselves.

If we expect the dark matter halo to rotate like a rigid body, it will give rise to a vortex lattice \cite{inv3,inv4}. It was already shown in laboratory experiments that such vortex lattices arise when an asymmetry is introduced \cite{madison2000vortex,madison2001stationary}. It has been shown experimentally that vortices  arise at a critical angular velocity, at which the energy of the system is lower if it generates vortices \cite{inv3}. When the angular velocity gets higher, instead of generating a bigger vortex, a lattice of several vortices will be generated\cite{madison2000vortex}.

How a vortex will influence the properties of the dark matter halo was studied in \cite{inv3}. Significantly, a core appears at the center of the vortex, and within the core the density is zero. In \cite{inv4}, this feature was used to explain the wiggles in the rotation curves of the galaxies.

For simplicity, we will not study the vortices in this paper. We will assume that the halo rotates like a rigid body, we will ignore the cores that generate inhomogeneities in the density profile, and we will concentrate our attention on how rotation causes the deformation of the halo,  and influences its observable physical properties.

\section{Deformation of the slowly rotating BEC dark matter halos}\label{sect2}

There have been many studies considering the problem of rotating polytropes, using different approximations and building different models, like, for example, in  \cite{cunningham1977rapidly,kong2015exact}. For a detailed discussion of the rotational properties of $n=1$ polytropes see \cite{Horedt}. We will assume the halo to be rotating slowly, and thus a first order approximation is sufficient. Chandrasekhar has worked on this problem in 1933 \cite{chandrasekhar1933equilibrium,ch33_2}, and we will mainly follow his method (for a comparative study of the different approaches to the rotation problem see \cite{kong2015exact}).

\subsection{The energy density and the gravitational potential of the rotating BEC dark matter halos}

We have already obtained the Helmholtz equation (\ref{Helm}) describing the matter distribution inside a rotating BEC dark matter halo. Its general solution is given by
\begin{equation}\label{2}
  \rho(r,\theta)=\frac{\omega^2}{2\pi G}+\sum_{l=0}^{\infty}A_{2l}j_{2l}(kr)P_{2l}(\cos{\theta}),
\end{equation}
where $j_l(X)$ are spherical Bessel functions, which are the solutions to the radial part of the equation, while $P_l(\cos{\theta})$ are the Legendre polynomials - they are the solutions to the angular part of the Helmholtz equation. In the solution, we have neglected all the terms of odd order, since it has been shown that a rotating mass must be symmetric about its equator. This result is called Lichtenstein's theorem \cite{lebovitz1967rotating}.
To determine the coefficients $A_{2l}$ in the first order of approximation, we will write down the solution for the gravitational potential, and use the fact that it is continuous at the boundary of the dark matter halo.

The gravitational potential satisfies the equations
\begin{equation}
\nabla^2V=4\pi G\rho, r\leq r_0,
\ee
\be
\nabla^2V=0, r\geq r_0,
\end{equation}
where $r_0$ is the radius (boundary) of the halo.
We can represent the potential in a general form as
\begin{equation}\label{1.2}
  V=V_0(kr)+\sum_{l=1}^{\infty}V_{2l}(kr)P_{2l}(\cos{\theta}).
\end{equation}
Then the gravitational potential equation inside the halo becomes
\bea\label{eqp1}
 && \frac1{r^2}\frac{\partial}{\partial r}\left(r^2\frac{\partial V}{\partial r}\right)+\frac1{r^2\sin\theta}\frac{\partial}{\partial\theta}\left(\sin\theta\frac{\partial V}{\partial\theta}\right)=\nonumber\\
 && 4\pi G\Bigg[\frac{\omega^2}{2\pi G}+\sum_{l=0}^{\infty}A_{2l}j_{2l}(kr)P_{2l}(\cos{\theta})\Bigg].
\eea
In the following we denote $\xi=kr$, and $\mu=\cos\theta$, respectively. Thus Eq.~(\ref{eqp1}) takes the form
\bea
&&  \frac1{\xi^2}\frac{\partial}{\partial\xi}\left(\xi^2\frac{\partial V}{\partial\xi}\right)+\frac1{\xi^2}\frac{\partial}{\partial\mu}\left[(1-\mu^2)\frac{\partial V}{\partial\mu}\right]=\nonumber\\
&&\frac{4\pi G}{k^2}\Bigg[\frac{\omega^2}{2\pi G}+\sum_{l=0}^{\infty}A_{2l}j_{2l}(\xi)P_{2l}(\mu)\Bigg].
\eea
Now we separate the terms in the above equation. For the 0th order term we obtain
\begin{equation}
  \frac1{\xi^2}\frac{\mathrm{d}}{\mathrm{d}\xi}\left(\xi^2\frac{\mathrm{d}V_0}{\mathrm{d}\xi}\right)=\frac{2\omega^2}{k^2}+\frac{4\pi G}{k^2}A_0j_0(\xi).
 \end{equation}
The solution of the above equation is
\begin{equation}
  V_0=\frac{\omega^2}{3k^2}\xi^2-\frac{4\pi G}{k^2}A_0j_0(\xi)+\mathrm{constant}.
\end{equation}
For the higher order terms, since the $P_{2l}$'s satisfy the equation
\begin{equation}
  \frac{\mathrm{d}}{\mathrm{d}\mu}\left[\left(1-\mu^2\right)\frac{\mathrm{d}P_j}{\mathrm{d}\mu}\right]+j(j+1)P_j=0,j\in N,
\end{equation}
we obtain the equations
\begin{equation}\label{32}
\frac1{\xi^2}\frac {\mathrm{d}}{\mathrm{d}\xi}\left(\xi^2\frac{\mathrm{d}V_{2l}}{\mathrm{d}\xi}\right)=\frac{2l(2l+1)}{\xi^2}V_{2l}+\frac{4\pi G}{k^2}A_{2l}j_{2l}(\xi).
\end{equation}
Since $\nabla^2(j_{2l}P_{2l})=-j_{2l}P_{2l}$, and by taking into account  that the Bessel functions $j_{2l}$ satisfy the equation
\be
  \frac1{\xi^2}\frac {\mathrm{d}}{\mathrm{d}\xi}\left(\xi^2\frac{\mathrm{d}j_{2l}}{\mathrm{d}\xi}\right)=\frac{2l(2l+1)}{\xi^2}j_{2l}-j_{2l},
  \ee
  it follows that
  \be
  j_{2l}=-\frac1{\xi^2}\frac{\mathrm{d}}{\mathrm{d}\xi}\left(\xi^2\frac{\mathrm{d}j_{2l}}{\mathrm{d}\xi}\right)+\frac{2l(2l+1)}{\xi^2}j_{2l}.
\ee
By substituting the above relations into Eq.~(\ref{32}) we obtain
\bea
 && \frac1{\xi^2}\frac{\mathrm{d}}{\mathrm{d}\xi}\left(\xi^2\frac{\mathrm{d}V_{2l}}{\mathrm{d}\xi}\right)-\frac{2l(2l+1)}{\xi^2}V_{2l}=\nonumber\\
&&  -\frac{4\pi G}{k^2}A_{2l}\left[\frac1{\xi^2}\frac{\mathrm{d}}{\mathrm{d}\xi}\left(\xi^2\frac{\mathrm{d}j_{2l}}{\mathrm{d}\xi}\right)-\frac{2l(2l+1)}{\xi^2}j_{2l}\right].
\eea
A particular solution of the above equation is
\begin{equation}
  V_{2l}=-\frac{4\pi G}{k^2}A_{2l}j_{2l}(\xi)+\mathrm{constant}.
\end{equation}
The regular solution of the equation
\be
\frac1{\xi^2}\frac{\mathrm{d}}{\mathrm{d}\xi}\left(\xi^2\frac{\mathrm{d}V_{2l}}{\mathrm{d}\xi}\right)-\frac{2l(2l+1)}{\xi^2}V_{2l}=0,
 \ee
 which behaves well near $\xi=0$ is
\begin{equation}
  V_{2l}=C_{2l}\xi^{2l},
\end{equation}
where $C_{2l}$ are arbitrary integration constants. Hence the general solution for the $V_{2l}$'s is
\begin{equation}
  V_{2l}=-\frac{4\pi G}{k^2}\left[A_{2l}j_{2l}(\xi)+B_{2l}\xi^{2l}\right]+\mathrm{constant},
\end{equation}
giving for the gravitational potential $V$ the general solution
\bea
  && V(\xi, \mu)=V_0(\xi)+\sum_{l=1}^{\infty}V_{2l}(\xi)P_{2l}(\mu) =\frac{\omega^2}{3k^2}\xi^2-\frac{4\pi G}{k^2}\times \nonumber\\
 &&\sum_{l=0}^{\infty}\left[A_{2l}j_{2l}(\xi)+B_{2l}\xi^{2l}\right]P_{2l}(\mu)+\mathrm{constant}=\nonumber\\
 && \frac{\omega^2}{3k^2}\xi^2-\frac{4\pi G}{k^2}\left[\rho-\frac{\omega^2}{2\pi G}+\sum_{l=0}^{\infty}B_{2l}\xi^{2l}P_{2l}(\mu)\right]+\nonumber\\
 &&\mathrm{constant}.
\eea

\subsection{The continuity conditions}

To determine the $B_{2l}$'s, we consider the behavior of the pressure $P$, which is given by the static limit of the Euler equation (\ref{Euler}) as
\begin{equation}
 \frac{\partial P}{\partial r}=-\rho\frac{\partial V}{\partial r}+\rho\omega^2r(1-\mu^2),
 \ee
 or, equivalently,
 \be
 \frac{\partial P}{\partial\xi}=-\rho\frac{\partial V}{\partial\xi}+\frac23\frac{\rho\omega^2\xi}{k^2}\left[1-P_2(\mu)\right].
\end{equation}
After substituting the pressure with the use of the Bose-Einstein Condensate equation of state, and equating the coefficients of the series expansions, we obtain
\be
B_2=\frac{\omega^2}{12\pi G}, B_{2l}=0,\  l\neq1.
\ee
Therefore
\bea
  V(\xi,\mu)&=&\frac{\omega^2}{3k^2}\xi^2-\frac{4\pi G}{k^2}\sum_{l=0}^{\infty}A_{2l}j_{2l}(\xi)P_{2l}(\mu)-\nonumber\\
 && \frac{\omega^2}{3k^2}\xi^2P_2(\mu)+\mathrm{constant}.
\eea
To the accuracy we are working, this potential should be continuous with the external potential on the sphere of radius $\xi_1=kR$, where $R$ is the boundary of the non-rotating sphere,
\begin{equation}\label{V_e}
  V_e=\frac{4\pi G}{k^2}\sum_{l=0}^{\infty}\frac{C_{2l}}{\xi^{2l+1}}P_{2l}(\cos{\theta})+\mathrm{Constant}.
\end{equation}
To determine the coefficient $A_{2}$, we will equate the potentials and their first derivatives at $\xi_1$,
\begin{equation}
\begin{cases}
-A_2j_2(\xi_1)-\frac{\omega^2}{12\pi G}\xi_1^2=\frac{C_2}{\xi_1^3},\\
-A_2j_2'(\xi_1)-\frac{\omega^2}{6\pi G}\xi_1=-\frac{3C_2}{\xi_1^4}.
\end{cases}
\end{equation}
Hence we obtain
\begin{equation}
  A_2=-\frac{5\omega^2\xi_1^2}{12\pi G\left[3j_2(\xi_1)+j_2'(\xi_1)\xi_1\right]},
\end{equation}
and for all the $l\neq0,1$, $A_{2l}=0$.
$A_0$ is not determined by the boundary condition of potential at radius $R$, since an arbitrary constant can always be added to the gravitational potential. We can determine it by the boundary condition of the matter density at the center of the dark matter halo,
\begin{equation}
\frac{d\rho_0(0)}{d\xi}=0,
\rho_0(0)=\frac{\omega^2}{2\pi G}+A_0=\rho_c.
\end{equation}
Hence we find
\begin{equation}
  A_0=\rho_c-\frac{\omega^2}{2\pi G}.
\end{equation}

\subsection{The first order corrections to density and radius}

With the use of the expressions for the coefficients obtained above, after substituting  $\xi_1=\pi$ and $\xi =kr$, respectively, we find the first-order correction to the density $\rho $ of the rotating Bose-Einstein Condensate dark matter halo as
\begin{equation}\label{rho}
  \rho(r,\theta)=\frac{\omega^2}{2\pi G}+\left(\rho_c-\frac{\omega^2}{2\pi G}\right)j_0(kr)-\frac{5\omega^2\pi}{12G}j_2(kr)P_2(\cos{\theta}),
\end{equation}
or, equivalently,
\be\label{rhonew}
\frac{\rho(r,\theta)}{\rho _c}=\left(1-\Omega ^2\right)j_0(kr)+\Omega ^2\left[1-\frac{5\pi ^2}{6}j_2(kr)P_2(\cos{\theta})\right],
\ee
where
\be
\Omega ^2=\frac{\omega^2}{2\pi G\rho _c}=0.238\times \left(\frac{\omega }{10^{-16}\;{\rm s}^{-1}}\right)^2\times \left(\frac{\rho _c}{10^{-25}\;{\rm g/cm^3}}\right)^{-1}.
\ee

We will describe the effect of the rotation on the structure of the dark matter halo by using the dimensionless parameter $\Omega ^2$. In particular, the case of slow rotation corresponds to values of $\Omega ^2$ so that $\Omega ^2<<1$. In the limiting case $\Omega ^2\rightarrow 0$ we recover from Eq.~(\ref{rhonew}) the static limit for the halo density, $\rho(r)\rho _c=j_0(kr)$. The angular momentum is usually described by using the dimensionless spin parameter $\lambda =J|E|^{1/2}/GM^{5/2}$, where $J$ is the angular momentum, and $E$ is the gravitational energy of the halo \cite{rotgal}.  From a physical point of view the spin parameter $\lambda $ is the ratio of the actual angular
momentum of the galaxy and of the maximum angular momentum value needed for rotational support. Simulations have shown that $\lambda $ is in the range $0.02-0.10$ \cite{rotgal}.

\begin{figure}[!htb]
\centering
\includegraphics[scale=0.65]{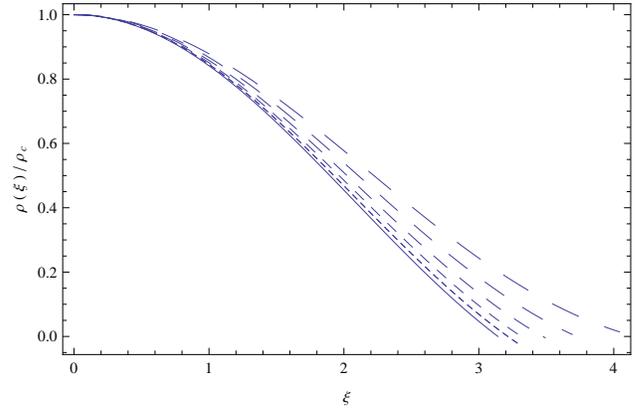}
\caption{Variation of the dimensionless ratio $\rho (\xi)/\rho _c$ as a function of $\xi =kr$ for a rotating Bose-Einstein Condensate dark matter halo for different values of $\Omega ^2$: $\Omega ^2=0$ (solid curve), $\Omega ^2=0.01$ (dotted curve), $\Omega ^2=0.0225$ (short dashed curve), $\Omega ^2=0.04$ (dashed curve), $\Omega ^2=0.0625$ (long dashed curve), and $\Omega ^2=0.09$ (ultra-long dashed curve), respectively.}\label{fig1}
\end{figure}

The comparative variation of the density in the  non-rotating and rotating cases at $\theta=\pi/2$ (corresponding to the equatorial plane) is presented in Fig.~\ref{fig1}.

To see how the halo is deformed by rotation, we need to obtain the boundary, or the radius where $\rho=0$.
Since the deformation from the spherical shape is small, in the first approximation we  write
\be
  \xi_0=\xi_1+\xi', \xi '<<\xi _1.
\ee
Expanding Eq.~(\ref{rho}) to the first order we have
\bea
 && \frac{\omega^2}{2\pi G}+\left(\rho_c-\frac{\omega^2}{2\pi G}\right)\left[j_0(\xi_1)+j_0'(\xi_1)\xi'\right]-\nonumber\\
&&  \frac{5\omega^2\pi}{12G}P_2(\cos{\theta})\left[j_2(\xi_1)+j_2'(\xi_1)\xi'\right]=0.
\eea
By substituting $\xi_1=\pi$  we obtain
\begin{equation}
  \xi'=\frac{3\left[2-5P_2(\cos\theta)\right]\omega^2}{12\pi G\rho_c+\left[-6+5P_2(\cos\theta)\right]\left(\pi^2-9\right)\omega^2}.
\end{equation}
In order to obtain the first order approximation we will further expand this expression to the first order of $\omega^2/2\pi G$, and obtain
\begin{equation}\label{xi}
  \xi'=\frac{2\omega^2-5\omega^2P_2(\cos{\theta})}{4G\rho_c}.
\end{equation}
This has the same form as in Chandrasekhar's work \cite{chandrasekhar1933equilibrium}. Hence the boundary of the slowly rotating Bose-Einstein Condensate dark matter halo is located at
\bea\label{r_0}
\hspace{-0.6cm}  r_0(\theta)&=&\frac{\pi}k+\frac{\omega^2}{4G\rho_c k}\left[2-5P_2(\cos{\theta})\right]=\frac{\pi}{k}\times \nonumber\\
\hspace{-0.6cm} && \left\{1+\Omega ^2\left[1-\frac{5}{2}P_2(\cos \theta)\right]\right\}.
\eea
For the equatorial radius of the dark matter halo we obtain
\be
r_0\left(\frac{\pi}{2}\right)= \frac{\pi}{k}\left(1+\frac{9}{4}\Omega ^2\right).
\ee
In the the non-rotating case only the first term exists. Rotation adds an expansion and an ellipticity term to the halo radius.

\section{Gravitational and astrophysical properties  of rotating Bose-Einstein Condensate dark matter halos}\label{sect3}

In the present Section we will obtain some basic gravitational and astrophysical properties of the slowly rotating Bose-Einstein Condensate dark matter halos, which could allow an in depth comparison of the theoretical model with the astronomical observations. In particular, we will consider the mass distribution within the halo, as well as to its gravitational potential energy. Moreover, we will concentrate on astrophysical parameters like velocity dispersion, logarithmic density and velocity slopes, and the tangential velocity expression, which allows a detailed comparison of the model predictions with observational data.

\subsection{Mass and gravitational potential of the slowly rotating BEC halo}

As we have already seen, the general solution for the matter density distribution $\rho$ inside the rotating halo in spherical coordinates is given by
\begin{equation}
  \rho(r,\theta)=\frac{\omega^2}{2\pi G}+\sum_{l=0}^{\infty}A_{2l}j_{2l}(kr)P_{2l}(\cos{\theta}).
\end{equation}
The boundary (radius) $r_0$ of the halo is defined as the surface whose points satisfy the condition $\rho(r_0)=0$. The equation for the density involves infinitely many terms in general.

The mass profile $m(r,\theta=\pi)$ within a radius $r$ is given by
\begin{equation}
m\left(r,\theta=\pi\right) =2\pi\int_0^{\pi}\sin{\theta}\mathrm d \theta\int_0^r r^2\rho(r,\theta)\mathrm d r,
\end{equation}
\\
which involves an integration of the spherical Bessel functions, which can be done by using the relation
\bea
&&\int_0^{r_0}j_{2l}(kr)r^2\mathrm{d}r
=\sqrt{\pi } 2^{-2 (l+1)} r_0^3 \Gamma \left(l+\frac{3}{2}\right) (kr_0)^{2 l}\times \nonumber\\
&& \, _1\tilde{F}_2\left(l+\frac{3}{2};l+\frac{5}{2},2 l+\frac{3}{2};-\frac{1}{4} k^2r_0^2\right),
\eea
\\
where $\,  _1\tilde{F}_2$ is the is the regularized generalized hypergeometric function $_pF_q\left(a;b;z\right)/\left(\Gamma \left(b_1\right)...\Gamma \left(b_q\right)\right)$.
Thus we obtain for the total mass the expression
\begin{widetext}
\bea
 M\left(r_0\right)&=&\int_0^{\pi}\frac{\omega^2r_0^3}{3G}\sin\theta\mathrm{d}\theta+
 2\pi^{3/2}\sum_{l=0}^{\infty}A_{2l}\int_0^{\pi}\mathrm{d}\theta P_{2l}(\cos\theta)\sin\theta \times \nonumber\\
 &&\Bigg[\sqrt{\pi}2^{-2(l+1)}r_0^3\Gamma \left(l+\frac{3}{2}\right) (kr_0)^{2 l} \,  _1\tilde{F}_2\left(l+\frac{3}{2};l+\frac{5}{2},2 l+\frac{3}{2};-\frac{1}{4} k^2r_0^2\right)\Bigg].
 \eea
 \end{widetext}

\begin{figure}[!htb]
\centering
\includegraphics[scale=0.65]{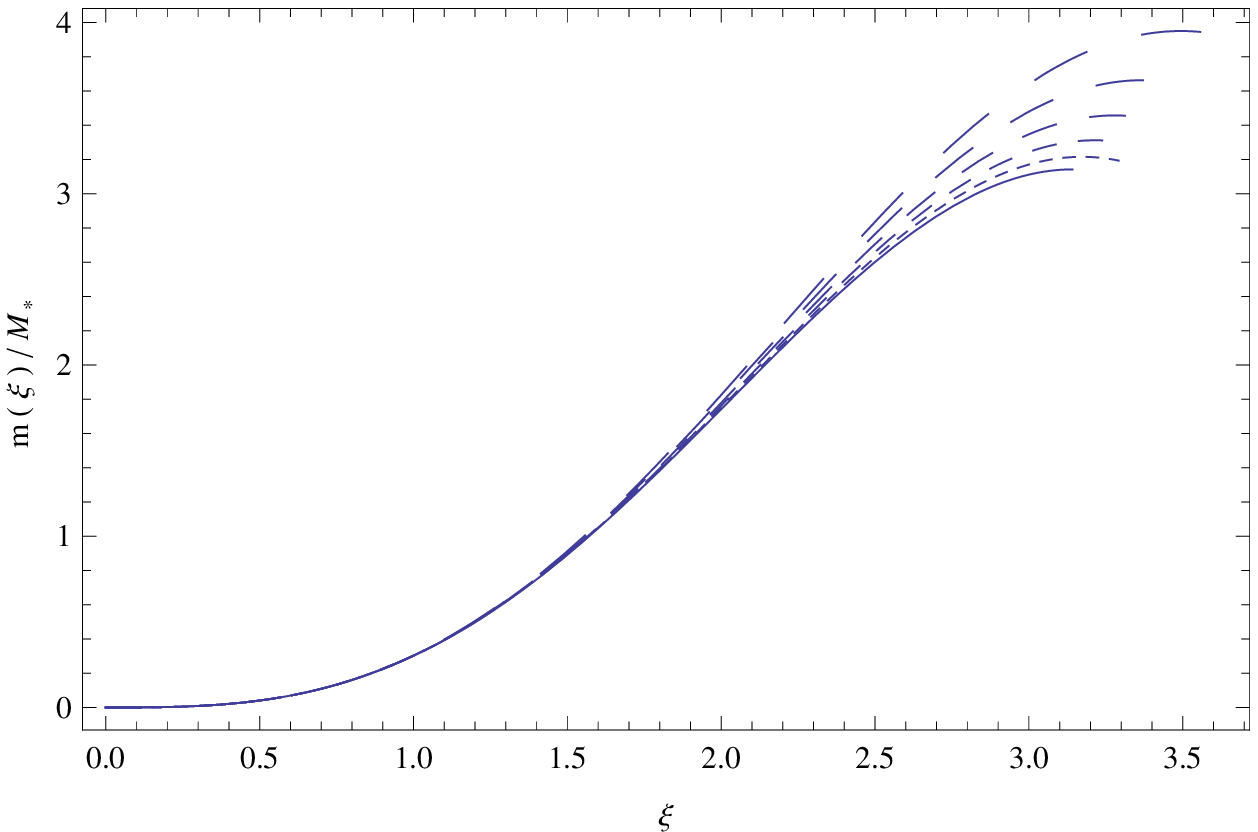}
\caption{Variation of the dimensionless ratio $m(\xi) /M _{*}$, $M_{*}=4\pi \rho _c/k^3$, as a function of $\xi =kr$ for a rotating Bose-Einstein Condensate dark matter halo for different values of $\Omega ^2$: $\Omega ^2=0$ (solid curve), $\Omega ^2=0.01$ (dotted curve), $\Omega ^2=0.0225$ (short dashed curve), $\Omega ^2=0.04$ (dashed curve), $\Omega ^2=0.0625$ (long dashed curve), and $\Omega ^2=0.09$ (ultra-long dashed curve), respectively.}\label{fig2}
\end{figure}

For dark matter halos located within a radius $r\leq \pi /k-3\omega^2/4G\rho_c k$, we can calculate the mass distribution within radius $r$ as being given by
\bea
m(r)&=&2\pi\int_0^{\pi}\sin{\theta}\mathrm d \theta\int_0^r r^2\rho(r,\theta)\mathrm d r=\nonumber\\
&&-\frac{4\pi\rho_c}{k^3}(kr\cos{kr}-\sin{kr})+\frac{2\omega^2}{k^3G}\times \nonumber\\
&&\left[\frac{(kr)^3}3+kr\cos{kr}-\sin{kr}\right]=\frac{4\pi \rho _c}{k^3}(kr)\times \nonumber\\
&&\Bigg[\left(1-\Omega ^2\right)\frac{\sin kr}{kr}-\left(1-\Omega ^2\right)\cos kr +\Omega ^2\frac{(kr)^2}{3}\Bigg]. \nonumber\\
\eea
As compared to the non-rotating case, a second term appears, which is due to the presence of the rigid body type rotation. The mass profile within radius $r$ is bigger than in the non-rotating case, and it depends on the central density  $\rho_c$. These results are consistent with the slower decay of the density profile for larger values of the radial coordinate $r$. The variation of the dimensionless ratio $m(\xi)/M_{*}$, where $M_{*}=4\pi \rho _c/k^3$ is represented in Fig.~\ref{fig2}.

The total mass of the condensate in the first order of $\omega^2/2\pi G$ is
\bea\label{64}
M\left(r_0\right)&=&2\pi\int_0^{\pi}\sin{\theta}\mathrm d \theta\int_0^{r_0} r^2\rho(r,\theta)\mathrm d r \simeq \frac{4\pi^2\rho_c}{k^3}+\nonumber\\
 &&\frac{2\pi\omega^2}{Gk^3}\left(\frac{\pi^2}3-1\right)=\frac{4\pi ^2}{k^3}\rho _c\left[1+\left(\frac{\pi ^2}{3}-1\right)\Omega ^2\right].\nonumber\\
\eea
For a given galaxy,  the total
halo mass is fixed by the physical processes leading to its formation, and it is unchanged due to the presence of rotation. But, as one can see from Eq.~(\ref{64}), a rotating halo is able to hold more mass than a static one, and in this sense the rotation of the dark matter halo becomes a stabilizing factor against gravitational collapse.

To perform the integration in the accuracy of the first order of $\omega^2/2\pi G$, we first integrate over $r$, then expand the result in the first order, and then we perform the integration over $\mu$. This procedure does not affect the final result obtained by doing the series expansion after performing the full integration, since $\omega^2/2\pi G$ is independent of $r$ and $\mu$, and thus the power of it is not changed after each integration.

Then we can express the central density $\rho_c$ as a function of the total mass and angular velocity as
\begin{equation}
  \rho_c=\frac{\pi}{4R^3}\left[M-\frac{2\pi\omega^2}{k^3}\left(\frac{\pi^2}3-1\right)\right].
\end{equation}
The total volume of the halo becomes
\bea
V_{BEC}&=&2\pi\int_0^{\pi}\sin{\theta}\mathrm d \mu\int_0^{r_0}r^2\mathrm d r
\simeq \frac43\pi R^3+\frac{2R^3\omega^2}{G\rho_c}=\nonumber\\
&&\frac43\pi R^3\left(1+\Omega ^2\right).
\eea
From the above expression of the volume it follows that due to rotation the halo has expanded. The added volume is proportional to $\omega^2$, and inversely proportional to $\rho_c$.

With the help of the total mass and of the volume we obtain the mean density of the BEC halo as
\bea
\langle\rho\rangle &=&\frac M{V_{BE}} \simeq\frac{3\rho_c}{\pi^2}+\left(1-\frac{12}{\pi^2}\right)\frac{\omega^2}{2\pi G}=\nonumber\\
&&\frac{3\rho _c}{\pi ^2}\left[1+\frac{\pi ^2}{3}\left(1-\frac{12}{\pi ^2}\right)\Omega ^2\right]<\frac{3\rho_c}{\pi^2},
\eea
which is smaller than in the non-rotating case.

In the previous Section, we have already calculated the gravitational potential inside the halo,
\bea
V(r,\theta)&=&\frac{\omega^2}{3k^2}(kr)^2-\frac{4\pi G}{k^2}\Bigg[\left(\rho_c-\frac{\omega^2}{2\pi G}\right)j_0(kr)-\nonumber\\
&&\frac{5\omega^2\pi}{12G}j_2(kr)P_2(\cos\theta)\Bigg]
  -\frac{\omega^2}{3k^2}(kr)^2P_2(\cos\theta)+\nonumber\\
 && \mathrm{constant}
\eea
We can determine the constant by using the continuity of the potential near radius $\xi _1=kR$,
\bea
&&\frac{\omega^2}{3k^2}\xi_1^2-\frac{4\pi G}{k^2}(\rho_c-\frac{\omega^2}{2\pi G})j_0(\xi_1)+\mathrm{constant}=\frac{4\pi G}{k^2}\frac{C_0}{\xi_1},\nonumber\\
&&\frac{2\omega^2}{3k^2}\xi_1-\frac{4\pi G}{k^2}(\rho_c-\frac{\omega^2}{2\pi G})j_0'(\xi_1)=-\frac{4\pi G}{k^2}\frac{C_0}{\xi_1^2}.
\eea
Thus we obtain
\begin{equation}
  \mathrm{constant}=-\frac{4\pi G}{k^2}\rho_c-\frac{\omega^2}{k^2}(\pi^2-2).
\end{equation}
Hence the gravitational potential is given by
\bea\label{V}
&&  V(r,\theta)=-\frac{4\pi G}{k^2}\rho_c\left[1+j_0(kr)\right]+\frac{2\omega^2}{k^2}j_0(kr)-\nonumber\\
&&\frac{5\omega^2\pi^2}{3k^2}j_2(kr)P_2(\cos{\theta})
  +\frac{\omega^2}{3k^2}(kr)^2\left[1-P_2(\cos{\theta})\right]-\nonumber\\
&&  \frac{\omega^2}{k^2}\left(\pi^2-2\right).
\eea
Using the gravitational potential, we can calculate the gravitational binding energy $U$ defined as
\begin{equation}\label{U_g}
  U(r)=\frac12\int \rho(r,\theta)V(r,\theta)\mathrm d^3\boldsymbol{r}.
\end{equation}
Hence the total gravitational potential energy of the BEC dark matter halo is given by
\begin{equation}
\begin{aligned}
U(r_0)=&\pi\int_0^{\pi}\sin{\theta}\mathrm d \theta\int_0^{r_0}r^2\rho(r,\theta)V(r,\theta)\mathrm d r
\\ \simeq &-\frac{12\pi^3G\rho_c^2}{k^5}+\frac{4\pi^2\rho_c\omega^2}{k^5}\left(1-\frac23\pi^2\right).
\end{aligned}
\end{equation}
As compared to the non-rotating case, the total gravitational energy has a second negative term, and hence for certain values of $\rho_c$ it is lower than in the non-rotating case.

The centrifugal potential of the rotating dark matter halo is
\begin{equation}
V_{cen}=-\frac12\omega^2(x^2+y^2)
=-\frac12\omega^2r^2\sin^2{\theta}.
\end{equation}
Hence we can calculate the centrifugal potential energy as
\bea
U_{cen}&=&2\pi\int_0^{\pi}\sin{\theta}\mathrm d \theta\int_0^{r_0}r^2\rho(r,\theta)V_{cen}\mathrm d r \nonumber\\
&& \simeq
 \frac{4\pi^2\rho_c\omega^2}{k^5}\left(2-\frac{\pi^2}3\right).
\eea
The centrifugal potential energy is always lower than $0$.

Hence the effective potential energy in the corotating frame can be obtained as
\begin{equation}\label{U}
  U_{eff}=U+U_{cen}=\frac{12\pi^3G\rho_c^2}{k^5}\left[1+\frac{2}{3}\left(\pi^2-3\right)\Omega ^2\right].
\end{equation}

\subsection{Velocity dispersion of particles in slowly rotating BEC halos}

In the following we consider the dynamics of a collection of particles (stars) in the gravitational field of a Bose-Einstein Condensate dark matter halo. The statistical properties of the motion are described by the Jeans equation, which for a system of particles with number density $n=n(x_i,t)$ is given by \cite{BT}
\bea
&&n\frac{\partial <v_j>}{\partial t}+\sum_i{n<v_i>\frac{\partial <v_j>}{\partial x_i}}=\nonumber\\
&&-n\frac{\partial V}{\partial x_j}-\sum _i{\frac{\partial \left(n\sigma _{ij}^2\right)}{\partial x_i}},
\eea
where the notation $<…>$ means an average at a given point and time $(x,t)$, and
\bea
  \sigma_{ij}^2&\equiv& <\left(v_i-<v_i>\right)\left(v_j-<v_j>\right)>=\nonumber\\
&&  <v_iv_j>-<v_i><v_j>.
\eea
 For a static spherical symmetrical system, one can further simplify the radial Jeans equation by adopting  the assumptions:
1) steady-state hydrodynamic equilibrium, which implies
$\frac{\partial v_j}{\partial t}=0$, and $<v_r>=0$, respectively,
2)
$<v_{\theta}>=<v_{\phi}>=0$, and $\sigma_{r\theta}^2=\sigma_{r\phi}^2=\sigma_{\theta\phi}^2=0$, respectively, which follows from the spherical symmetry of the system, and 3) a single tangential velocity dispersion for all directions $\sigma_{tt}^2=\sigma_{\theta\theta}^2=\sigma_{\phi\phi}^2$.

Hence when the tangential velocity dispersion tensor is isotropic, $\sigma_{rr}^2=\sigma_{tt}^2=\sigma^2$, the Jeans equation reduces to
\begin{equation}
  \frac1{\rho_n}\frac{\partial\rho_n\sigma^2}{\partial r}=-\frac{\partial V}{\partial r}.
\end{equation}
By assuming that all  particle have the same mass, the particle velocity dispersion is obtained as
\begin{equation}
  \sigma^2=\frac1{\rho}\int_r^{\infty}\rho\frac{\partial V}{\partial r}\mathrm{d}r.
\end{equation}

For a rotating system, we can still adopt the  following set of assumptions,
1) $\frac{\partial v_j}{\partial t}=0$, $\langle v_r\rangle=\langle v_{\theta}\rangle=0$,
2)  $\langle v_iv_j\rangle=0$, and
3) $\sigma_{rr}^2=\sigma_{\phi\phi}^2=\sigma^2$ (isotropic velocity distribution), respectively.

Thus in presence of rotation the radial Jeans equation becomes
\begin{equation}
  \frac{\partial}{\partial r}(\rho_n\sigma^2)-\frac{\rho_n}{r}\langle v_{r}^2\rangle=-\rho_n\frac{\partial V}{\partial r}.
\end{equation}\\
Therefore for the mean value of the square of the radial velocity we obtain
\bea
 &&\hspace{-1.2cm} \langle v_r^2(r,\theta)\rangle=\nonumber\\
&&\hspace{-1.2cm}\frac1{\rho(r,\theta)}\int_r^{r_0}\Bigg[\rho(r,\theta)\frac{\partial V(r,\theta)}{\partial r'}-\rho(r,\theta)\omega^2r'\sin^2\theta\Bigg]\mathrm{d}r'.
\eea
In order to find an explicit expression for $\langle v_r^2(r,\theta)\rangle$, we expand the integrand to the first order in $\omega^2/2\pi G$.  Hence for $r<r_0$ we find
\begin{widetext}
\bea
  \langle v_r^2(r,\theta)\rangle&=&\frac{2\rho_cG\pi}{k^3r}\sin kr+\frac{\omega^2}{24k^5r^3}\cos(kr)\Bigg\{-12k^2r^2+10P_2(\cos\theta)\Bigg[6k^4r^4+\pi^2(6-7k^2r^2)\Bigg]+\nonumber\\
&&  \Bigg[12k^2r^2+10P_2(\cos\theta)\pi^2(-6+k^2r^2)\Bigg]\cos(2kr)
  +24k^3r^3\sin(kr)+30kP_2(\cos\theta)\pi^2r\sin(2kr)\Bigg\}.
  \eea
  \end{widetext}
We define the kinetic energy $K$ of the halo in terms of the average velocity dispersion $\sigma_v$ as
\begin{equation}
  K(r)=\frac32\int\rho(r,\theta)\sigma_v^2(r,\theta)\mathrm{d}V.
\end{equation}
If the velocity dispersion is a constant,
\begin{equation}
  K(r)=\frac32M(r)\sigma_v^2,
\end{equation}
and at the boundary $K$ takes the value,
\begin{equation}
  K(r_0)=\frac{6\pi^2\rho_c}{k^3}\left[1+\Omega ^2\left(\frac{\pi^2}3-1\right)\right]\sigma_v^2.
\end{equation}
For the ratio of the kinetic and potential energy, after expanding to the first order of $\omega^2/2\pi G$,
we obtain
\bea
  \frac{K\left(r_0\right)}{|U\left(r_0\right)|}&\simeq & \Bigg[\frac{k^2}{2\pi G \rho_c}-\frac{k^2\left(\pi ^2+3\right)}{36\pi^2G^2\rho_c^2}\omega^2\Bigg]\sigma_v^2=\nonumber\\
 && \frac{k^2}{2\pi G \rho_c}\left(1-\frac{\pi ^2+3}{9}\Omega ^2\right)\sigma _v^2.
\eea
Comparing this expression to the non-rotating case, we can see that rotation generates a second (negative) term in the parentheses, and hence in the presence of rotation the ratio of the kinetic and potential energy of particles in motion in rotating BEC dark matter halo is lower than in the non-rotating case.

\subsection{The logarithmic density slopes}

The logarithmic density slope of the rotating dark matter halo is defined by \cite{Har1}
\be
\alpha_{BEC}(r,\theta)=\frac{d\ln\rho (r,\theta)}{d\ln r}.
\ee
By taking into account the general solution of the Helmholtz equation \ref{Helm} we can calculate the logarithmic slope of the rotating BEC dark matter halo as follows. First for the derivative of the density with respect to the radial dimensionless variable $kr$ we have (for the proof see Appendix~\ref{app})
\be
\frac{\mathrm{d}\rho}{\mathrm{d}(kr)}=\sum^{\infty}_{l=0}A_{2l}\Big[-\frac{2l+1}{kr}j_{2l}(kr)+j_{2l-1}(kr)\Big]P_{2l}(\cos\theta).
\ee

Hence for the logarithmic slope of the density we obtain after a simple calculation
\bea
&&\alpha_{BEC}(r)=\frac{r\mathrm{d}\rho}{\rho \mathrm{d}r}=\nonumber\\
&&\frac{kr\sum^{\infty}_{l=0}A_{2l}\Big[-\frac{2l+1}{kr}j_{2l}(kr)+j_{2l-1}(kr)\Big]P_{2l}(\cos\theta)}{\frac{\omega^2}{2\pi G}+\sum_{l=0}^{\infty}A_{2l}j_{2l}(kr)P_{2l}(\cos{\theta})}.\nonumber\\
\eea
By taking into account the explicit expression of the coefficients in the solution of the Helmholtz equation we finally find
\begin{widetext}
\bea\label{94}
\alpha_{BEC}(r,\theta)=\frac{\mathrm{d}\ln\rho (r,\theta)}{\mathrm{d}\ln r}
=-\frac{6\left(\rho_c-\frac{\omega^2}{2\pi G}\right)(kr\cos kr-\sin kr)-\frac{5k\pi r\omega^2}{2G}P_2(\cos\theta)\Big[kr j_1(kr)-3j_2(kr)\Big]}{kr\left\{6\left(-\rho_c+\frac{\omega^2}{2\pi G}\right)j_0(kr)+\frac{\omega^2}{2\pi G}\left[-6+5P_2(\cos\theta)\pi^2j_2(kr)\right]\right\}}.
\eea
\end{widetext}
By expanding Eq.~(\ref{94}) to the first order of $\omega^2/2\pi G$, we obtain
\bea\label{alpha}
 && \alpha_{BEC}(r,\theta)=-\left[1-kr\cot{kr}\right]-\frac{\omega^2}{12\pi G\rho_c}kr\cos(kr)\times \nonumber\\
  && \Bigg\{-6+6kr\cot{kr}+ 5P_2(\cos{\theta})\pi^2\Bigg[krj_1(kr)-\nonumber\\
 && \left(2+kr\cot{kr}\right)j_2(kr)\Bigg]\Bigg\}.
\eea
As compared to the non-rotating case, a third term is added to the expression of the logarithmic slope of the density. The third term is smaller than the the non-rotating value of $\alpha_{BEC}$, and varies with $\theta$. This seems reasonable, since the rotation pushes the matter outward, and the ratio of the center density to the density at larger radii is smaller.

Since the logarithmic density slope varies with $\theta$, it's no longer convenient to define the core radius $R_{core}$ as $\alpha_{BEC}(R_{core})=1$. If we simply define it as $R_{core}=nR=n\pi/k$, where $n$ is a constant which must be determined from observations ($n=0.6$ gives the value of the core radius in the static case \cite{Har1}), we can define the mean value of the logarithmic density slope within the radius $0\leq r\leq R_{core}$ as
\be
\langle\alpha_{BEC}\rangle=\frac1{R_{core}}\int_0^{R_{core}}\left.\alpha_{BEC}(r,\theta)\right|_{\theta =\pi /2}\mathrm{d}r.
\ee

\begin{figure}[!htb]
\centering
\includegraphics[scale=0.65]{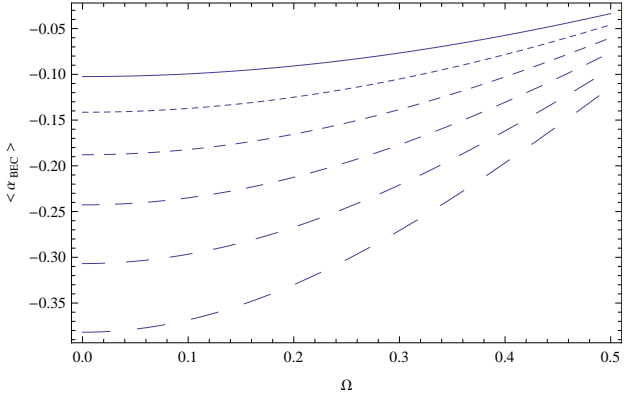}
\caption{Variation of the mean value of the logarithmic density slope $\langle\alpha_{BEC}\rangle$ for a rotating Bose-Einstein Condensate dark matter halo as a function of $\Omega$,  for different values of the core radius $R_{core}$: $R_{core}=0.30R$ (solid curve), $R_{core}=0.35R$ (dotted curve), $R_{core}=0.40R$ (short dashed curve), $R_{core}=0.45R$ (dashed curve), $R_{core}=0.50R$ (long dashed curve), and $R_{core}=0.55R$ (ultra-long dashed curve), respectively.}\label{fig3}
\end{figure}

The variation of the mean value of the logarithmic slope of the density $\langle\alpha_{BEC}\rangle$ is represented, as a function of $\Omega$, and for different values of $R_{core}$, in Fig.~\ref{fig3}.

Also, we can calculate the density at the core radius,
\bea
  \rho(R_{core})&=&\frac2{\pi}\rho_c \Bigg\{1+\frac{\pi}{2}\Omega ^2\Bigg[1-\frac2{\pi}+\nonumber\\
 && \frac{5P_2(\cos\theta)\left(-12+\pi^2\right)}{3\pi}\Bigg]\Bigg\}.
\eea
The core density is smaller as compared to the non-rotating case at $\theta=0$, and larger at $\theta=\pi/2$.

For a given $\rho_c$, the quantity  $\rho_cR_{core}$ can be obtained as
\begin{equation}
  \rho_cR_{core}=\frac{\pi}{8R^2}\left[M-\frac{2\pi\omega^2}{k^3}\left(\frac{\pi^2}3-1\right)\right],
\end{equation}
and its value depends on $\omega^2$.

\subsection{Tangential velocity of test particles in slowly rotating BEC ark matter halos}

In the Newtonian approximation the tangential velocity of a test particle moving in the Bose-Einstein condensed dark matter halo is given by
\begin{equation}
  V_{tg}^2(r)=\frac{Gm(r)}r.
\end{equation}

\begin{figure}[!htb]
\centering
\includegraphics[scale=0.65]{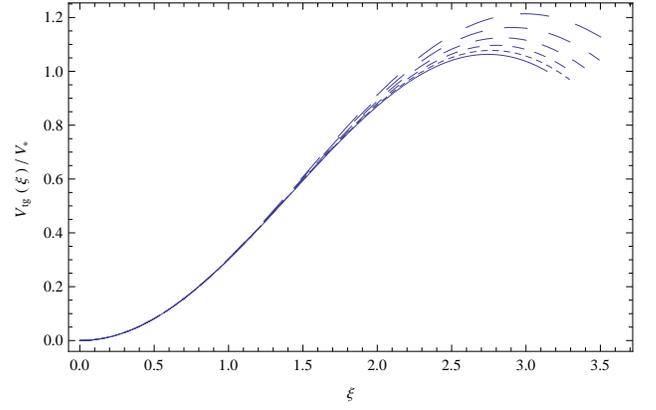}
\caption{Variation of the dimensionless tangential velocity of test particles $V_{tg}(\xi) /V _{*}$, $V_{*}=4\pi \rho _c/k^2$, as a function of $\xi =kr$, for a rotating Bose-Einstein Condensate dark matter halo for different values of $\Omega ^2$: $\Omega ^2=0$ (solid curve), $\Omega ^2=0.01$ (dotted curve), $\Omega ^2=0.0225$ (short dashed curve), $\Omega ^2=0.04$ (dashed curve), $\Omega ^2=0.0625$ (long dashed curve), and $\Omega ^2=0.09$ (ultra-long dashed curve), respectively.}\label{fig4}
\end{figure}

In the slow rotation approximation, and for $r\leq \pi/ k-3\omega^2/4G\rho_ck$, we have
\bea\label{vhalo}
\hspace{-0.6cm}  V_{tg}^2(r)&=&-\frac{4\pi G\rho_c}{k^2}\Bigg[\cos(kr)-\frac{\sin(kr)}{kr}\Bigg]+ \frac{2\omega^2}{k^2}\times \nonumber\\
\hspace{-0.6cm}&& \Bigg[\frac{(kr)^2}3+\cos kr-\frac{\sin kr}{kr}\Bigg]=\frac{4\pi G\rho _c}{k^2}\times \nonumber\\
\hspace{-0.6cm}&&\Bigg\{\left(1-\Omega ^2\right)\left[\frac{\sin kr}{kr}-\cos kr\right]+\frac{\Omega ^2}{3}(kr)^2\Bigg\}.
\eea
We can see that due to rotation a second positive term is added to the expression of the tangential velocity, so that at a distance $r$, the tangential velocity of a test particle becomes higher as compared to the non-rotating case. This is because the mass profile within the radius is higher. The variation of the ratio $V_{tg}(\xi)/V_{*}$ as a function of $\xi =kr$, where $V_{*}=4\pi G\rho_c/k^2$, is represented in Fig.~\ref{fig4}.

At the equatorial boundary of the halo we have,
\bea
  V^2_{tg}\left(r_0,\frac{\pi}2\right)&=&\frac{GM(r_0)}{r_0(\pi/2)}\simeq\frac{4\pi G\rho_c}{k^2}+\frac{4\pi^2-39}{6 k^2}\omega^2=\nonumber\\
&& \frac{4 G\rho_c R^2}{\pi}\left(1+\frac{4\pi ^2-39}{12}\Omega ^2\right) .
\eea
We can see that the second term proportional to $\omega ^2$ is also positive.

 Hence in a rotating BEC dark matter halo, the tangential velocity of a test particle is larger than in the static one. Since the ratio of the density at larger and smaller radii is greater than in the non-rotating case, it follows that the tangential velocity is bigger at larger radii than in the non-rotating case.

An important observational quantity is the logarithmic slope of the tangential velocity $\beta _{tg}$,  defined by
\begin{equation}
  \beta_{tg}(r)=-\frac{\mathrm{d}\ln V_{tg}(r)}{\mathrm{d}\ln r}.
\end{equation}
Generally,  $\beta_{tg}$ can be obtained as
 \bea
\beta_{tg}(r)&=&-\frac{d\ln V_{tg}}{d\ln r}
 =-\frac{d\ln V_{tg}^2}{2d\ln r}=-\frac{rdV_{tg}^2}{2V_{tg}^2dr}=\nonumber\\
&& -\frac r{2V_{tg}^2}\left[\frac{Gm'(r)}r-\frac{Gm(r)}{r^2}\right]=\nonumber\\
&&-\frac 12 \left[\frac{m'(r)}{m(r)} r-1\right].
\eea
At the center, where $r=0$, we have $\beta_{tg}(0)=-1$. This gives the same result as in the static case, since $\omega=0$ at the center.

\begin{figure}[!htb]
\centering
\includegraphics[scale=0.65]{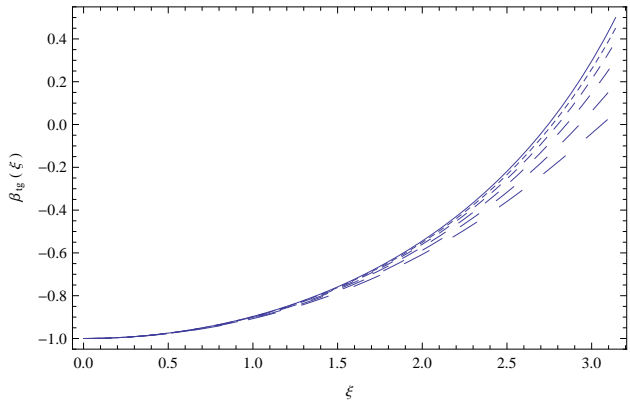}
\caption{Variation of the logarithmic slope of the tangential velocity $\beta _{tg}$ as a function of $\xi =kr$, for a rotating Bose-Einstein Condensate dark matter halo for different values of $\Omega ^2$: $\Omega ^2=0$ (solid curve), $\Omega ^2=0.01$ (dotted curve), $\Omega ^2=0.0225$ (short dashed curve), $\Omega ^2=0.04$ (dashed curve), $\Omega ^2=0.0625$ (long dashed curve), and $\Omega ^2=0.09$ (ultra-long dashed curve), respectively.}\label{fig5}
\end{figure}

By expanding the logarithmic slope of the tangential velocity to the first order in $\omega^2/2\pi G$, we obtain
\bea
  \beta_{tg}&\simeq& \frac12\left[1+\frac{k^2r^2}{kr\cot(kr)-1}\right]+\nonumber\\
 && \frac{k^3r^3\Bigg[3kr\cos(kr)+(-3+k^2r^2)\sin kr\Bigg]}{6(-kr\cos kr+\sin kr)^2}\Omega^2.\nonumber\\
\eea
This is smaller than the value without rotation. The variation of the logarithmic slope of the tangential velocity is shown, for different values of the dimensionless parameter $\Omega $, in Fig.~\ref{fig5}.

\section{Galactic rotation curves in the rotating Bose-Einstein Condensate dark matter model}\label{sect4}

 As a next step in our analysis we compare the predictions of the slowly rotating Bose-Einstein Condensate dark matter model with the observational data obtained for a sample of HSB, LSB and dwarf galaxies. From a realistic astrophysical point of view, the matter content in a galaxy consists of a distribution of baryonic (normal) matter, obtained as the algebraic sum of the masses $M_{star}$ of the stars, of the ionized gas with mass $M_{gas}$, of the neutral hydrogen of mass $M_{HI}$ etc., as well as of dark matter of mass $M_{DM}$. In the following we assume that dark matter is in the form of a slowly rotating Bose-Einstein Condensate. Therefore the total mass of the galaxy is can be obtained as $M_{tot} = M_{star} + M_{gas} + M_{HI} + M_{DM} + ... = M_{tot}^ B + M_{DM}$, where $M_{tot}^ B = M_{star} + M_{gas} + M_{HI} + ...$ is the total baryonic mass in the galaxy. The rotation velocity of a test particle $v_{rot}$ is given by the sum of the different matter contributions, as
 \be
 v_{rot}^2=v_{gas}^2+v_{stars}^2+...+v_{halo}^2,
 \ee
 where $v_{gas}^2$ and $v_{stars}^2$ are the contributions of the baryonic gas and stars, respectively, while $v_{halo}^2$ is the dark matter contribution, which we assume to be given by Eq.~(\ref{vhalo}). Hence the contribution of the rotating BEC dark matter halo to the rotational velocity can be represented as
 \\
% \begin{widetext}
 \bea\label{102}
 v_{halo}^2&=&80.563\times \left(\frac{\rho _c}{10^{-24}\;{\rm g/cm^3}}\right)\times \left(\frac{R}{{\rm kpc}}\right)^2\times \nonumber\\
 &&\Bigg\{\left[1-0.0238\left(\frac{\omega }{10^{-16}\;{\rm s}^{-1}}\right)^2\right]\times \nonumber\\
&&\left[\frac{\sin \left(\pi r/R\right)}{\pi r/R}-\cos \frac{\pi r}{R}\right]+0.00795\times \nonumber\\
 &&\left(\frac{\omega }{10^{-16}\;{\rm s}^{-1}}\right)^2\left(\frac{\pi r}{R}\right)^2\Bigg\}\;{\rm km}^2/{\rm s}^2.
 \eea
% \end{widetext}
 In Eq.~(\ref{102}), $R$ is {\it the radius of the static} BEC dark matter halo, which is fixed by the numerical values of the scattering length $a$ and the mass $m$ of the dark matter particle.

 \subsection{Fitting results}

 In order to test our model, we apply it to small nearby dwarf galaxies (size $\le 12$ kpc) and our Milky Way galaxy. We use the data of the Spitzer Photometry and Accurate Rotation Curves (SPARC) obtained in \cite{Lelli} for investigation. The baryonic components (bulge, disk and gas) included in the data are obtained by observations using the homogeneous surface photometry at 3.6 $\mu$m \cite{Lelli}. The rotation curve contributions of the baryonic components are only determined by the mass-to-luminosity ratios $\Upsilon \sim 1M_{\odot}/L_{\odot}$ of the disk and of the bulge. Nevertheless, the surface brightness and the resultant rotation curves obtained may also suffer from some systematic uncertainties, such as the irregularities in brightness profiles, uncertainties in inclination and patchy distribution of gas \cite{Lelli}. We choose the candidates which are Hubble stage $T=0-6$ galaxies, bulgeless galaxies, and with the distance to the galaxy $D \le 20$ Mpc. Generally speaking, these galaxies have less diffuse features and smaller uncertainties in the observational data. Based on these criteria, we consider 12 dwarf galaxies for testing the present Bose-Einstein Condensate dark matter model.

 Based on  Eq.~(\ref{102}), we have altogether four free parameters for fitting ($R$, $\rho_c$, $\omega$, $\Upsilon$). The mass-to-luminosity ratio of the disk $\Upsilon$ mainly affects the central rotation curve, while the other three control the entire shape of rotation curve. Here, we define the reduced $\chi^2$ as
 \be
 \chi_{\rm red}^2=\left(\frac{1}{N_{dof}}\right)\sum_i{\frac{\left(v_{rot,i}-v_{obs,i}\right)^2}{\sigma_i^2}},
 \ee
 \\
 where $N_{dof}$ is the number of degrees of freedom, $v_{rot,i}$ is the calculated rotation velocity, $v_{obs,i}$ is the observed rotation velocity, and $\sigma_i$ is the observational uncertainties of the rotation velocity. By minimizing the reduced $\chi^2$ value, we can obtain the best-fit values of $R$, $\rho_c$, $\omega$ and $\Upsilon$ for each of the considered dwarf galaxies (see Table~\ref{table1}).

 \begin{widetext}
\begin{center}
\begin{table}
\begin{tabular}{|l|c|c|c|c|c|}
  \hline
 Galaxy & $R$ (kpc) & $\rho_c$ $\left(10^{-24}\right)$ g cm$^{-3}$ & $\omega$ $\left(10^{-16}\right)$ s$^{-1}$ & $\Upsilon$ $\left(M_{\odot}/L_{\odot}\right)$ & $\chi_{\rm red}^2$ \\
  \hline
  NGC0024 & 8.0 & 0.9 & 3.2 & 1.8 & 0.96 \\
  NGC0100 & 6.9 & 1.1 & 2.9 & 0.5 & 0.38 \\
  NGC2976 & 8.2 & 4.0 & 0.1 & 0.5 & 0.51 \\
  NGC3877 & 6.0 & 6.0 & 2.9 & 0.1 & 1.81 \\
  NGC3949 & 7.6 & 1.4 & 5.0 & 0.5 & 0.55 \\
  NGC3972 & 4.9 & 2.4 & 3.6 & 0.6 & 1.15 \\
  NGC4051 & 6.7 & 1.6 & 2.7 & 0.5 & 0.37 \\
  NGC4085 & 4.9 & 4.4 & 3.1 & 0.2 & 1.34 \\
  NGC4389 & 7.9 & 1.6 & 4.0 & 0.1 & 0.41 \\
  UGC06667 & 4.4 & 2.4 & 3.1 & 0.7 & 0.87 \\
  UGC07151 & 2.9 & 3.2 & 3.3 & 0.7 & 1.01 \\
  UGC08286 & 7.0 & 0.7 & 3.0 & 2.5 & 2.83 \\
  \hline
 \end{tabular}
 \caption{The best-fit parameters for the 12 considered galaxies.}\label{table1}
\end{table}
\end{center}
\end{widetext}

\begin{figure*}[!htb]
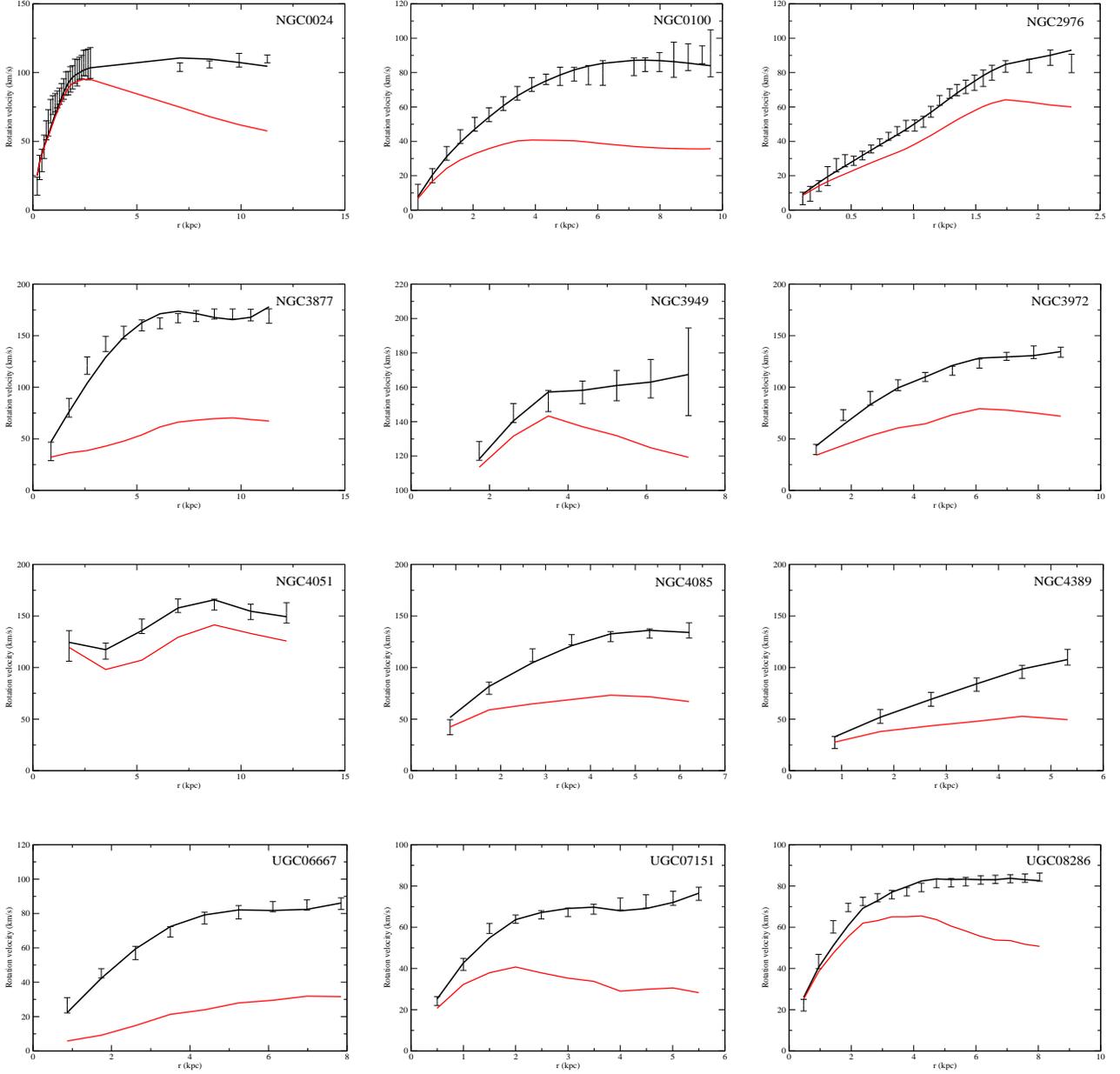

\centering
\includegraphics[width=150pt,height=100pt]{ngc0024.eps}\hspace{0.4cm}
\includegraphics[width=150pt,height=100pt]{ngc0100.eps}\hspace{0.4cm}
\includegraphics[width=150pt,height=100pt]{ngc2976.eps}\newline
\newline
\newline
\includegraphics[width=150pt,height=100pt]{ngc3877.eps}\hspace{0.4cm}
\includegraphics[width=150pt,height=100pt]{ngc3949.eps}\hspace{0.4cm}
\includegraphics[width=150pt,height=100pt]{ngc3972.eps}\newline
\newline
\newline
\includegraphics[width=150pt,height=100pt]{ngc4051.eps}\hspace{0.4cm}
\includegraphics[width=150pt,height=100pt]{ngc4085.eps}\hspace{0.4cm}
\includegraphics[width=150pt,height=100pt]{ngc4389.eps}\newline
\newline
\newline
\includegraphics[width=150pt,height=100pt]{ugc06667.eps}\hspace{0.4cm}
\includegraphics[width=150pt,height=100pt]{ugc07151.eps}\hspace{0.4cm}
\includegraphics[width=150pt,height=100pt]{ugc08286.eps}\newline
\caption{Best-fit rotational curves of the galaxy sample. The baryonic contributions are represented by the red lines (disk and gas contributions). The black lines represent the best-fit resultant rotation curves. The error bars are the observational data obtained in \cite{Lelli}.}
\label{fig6}
\end{figure*}

 In Fig.~\ref{fig6}, we show the best-fit rotation curves for the 12 dwarf galaxies.

 We can see that most of the best-fit values of $R$ fall into a small range $R=5-8$ kpc. Although the best-fit $R$ for the UGC07151 galaxy is quite small ($R=2.9$ kpc), the acceptable range of $R$ is $R=2.5-6.5$ kpc for $\chi_{\rm red}^2 \le 4$. Therefore, the results for the 12 dwarf galaxies are still consistent with a fixed value of $R \approx 6.5$.

 With the help of Eq.~(\ref{64}), which gives the mass of the Bose-Einstein Condensate dark halo, we can predict the mass of the considered sample of galaxies. In the limit of small rotational values we obtain for the total mass the expression
 \bea
 &&\hspace{-1cm}M=1.87043\times 10^7\times \left(\frac{ R}{{\rm kpc}}\right)^3 \times \frac{\rho _c}{10^{-24}\;{\rm g/cm^3}}\times \nonumber\\
 &&\hspace{-1cm}\left[1+0.0546393 \left(\frac{\omega}{10^{-16}\;{\rm s}^{-1}}\right)^2\left(\frac{\rho _c}{10^{-24}\;{\rm g/cm^3}}\right)^{-1}\right].\nonumber\\
 \eea

 The predicted masses of the 12 dwarf galaxies are presented in Table~\ref{table2}.

% \begin{widetext}
\begin{center}
\begin{table}
\begin{tabular}{|c|c|}
\hline
Galaxy & Mass ($10^{10}M_{\odot})$ \\
\hline
NGC0024 & 1.39 \\
\hline
NGC0100 & 0.95 \\
\hline
NGC2976 & 4.12 \\
\hline
NGC3877 & 2.60 \\
\hline
NGC3949 & 2.27 \\
\hline
NGC3972 & 0.68 \\
\hline
NGC4051 & 1.12 \\
\hline
NGC4085 & 1.08 \\
\hline
NGC4389 & 2.18 \\
\hline
UGC06667 & 0.46 \\
\hline
UGC07151 & 0.17 \\
\hline
UGC08286 & 0.76\\
\hline
\end{tabular}
 \caption{The predicted mass of the dark halo for the 12 considered galaxies.}\label{table2}
\end{table}
\end{center}
%\end{widetext}

We also apply our model to our Milky Way galaxy. By using the data obtained in \cite{Sofue} and the baryonic model used in \cite{Flynn}, we perform a similar fit for the Milky Way data (see Fig.~7). Since the mass-to-luminosity ratios of the bulge and disk are included in the baryonic model, we only have three free parameters to fit. The best-fit values are $R=6.9$ kpc, $\rho_c=3.0 \times 10^{-24}$ g cm$^{-3}$ and $\omega=2.6\times 10^{-16}$ s$^{-1}$ ($\chi_{\rm red}^2=0.54$). We also show the fit for $R=10$ kpc ($\chi_{\rm red}^2=1.02$) for reference in Fig.~7. Generally speaking, the best-fit values for the Milky Way galaxy and the dwarf galaxies are consistent with each other.

\begin{figure}[!htb]
\centering
\includegraphics[scale=0.35]{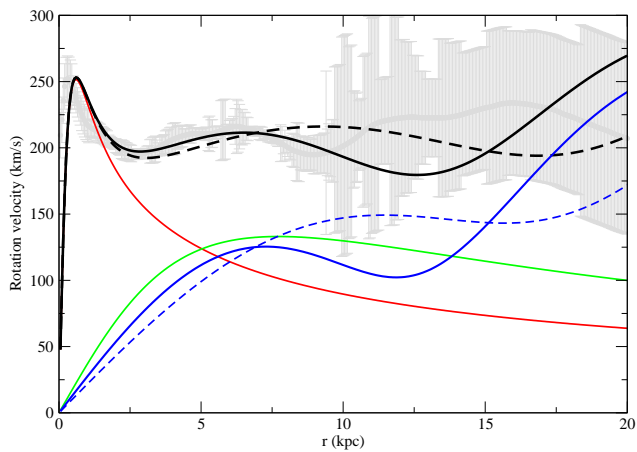}
\caption{The Milky Way rotation curve. The baryonic contributions are represented by the red line (bulge) and green line (disk). The blue lines are the dark matter contribution (solid: $R=6.9$ kpc; dashed: $R=10$ kpc). The black lines are the resultant rotation curves (solid: $R=6.9$ kpc; dashed: $R=10$ kpc). The grey lines and error bars are the observational data obtained in \cite{Sofue}.}\label{fig7}
\end{figure}

\section{Discussions and final remarks}\label{sect5}

 In the present paper we have investigated the possibility that the BEC dark matter halos could be in fact rotating. Presently, there is a common paradigm in the study of superfluidity according to which rotational motion in a Bose-Einstein condensate can exist only in the presence of
quantized vortices. However, in a recent numerical study \cite{Tsu} it was shown that the merging of two two-dimensional concentric Bose-Einstein
condensates with axial symmetry may lead to the formation of a
spiral dark soliton. This happens if one of the two condensates has a non-zero initial angular momentum. The spiral dark
soliton makes possible the transfer of angular momentum between the two condensates, and allows the merged
condensate to rotate, even in the absence of quantized vortices. A similar physical process could  have acted on a galactic scale, since galactic collisions and merging, which occurred frequently in the early stages of the evolution of the Universe, may have favored angular momentum transfer between galaxies. Therefore the rotation of BEC dark matter halos may be possible even in the absence of quantized vortices.

By using the Thomas-Fermi approximation, the basic evolution equations of a rotating BEC halo reduce to the case of the rotating polytropic spheres with index $n=1$. Many mathematical methods have been developed for the investigation of such systems. For example, in \cite{kong2015exact} the first exact analytic solution for an arbitrarily rotating gaseous polytrope under the assumption of the oblate spheroidal shape was derived. To obtain the solution the authors have adopted oblate spheroidal coordinates $\left(\xi, \eta, \phi\right)$, which are related to the Cartesian coordinates by means of the transformation $x=f\sqrt{\left(1+\xi ^2\right)\left(1-\eta ^2\right)}\cos \phi$, $x=f\sqrt{\left(1+\xi ^2\right)\left(1-\eta ^2\right)}\sin \phi$, $z=f\xi \eta$, where $f=\sqrt{R_e^2-R_p^2}/R_e$, where $R_e$ and $R_p$ are the equatorial and polar radii, respectively. In \cite{cunningham1977rapidly} index $n=1$ polytropes were studied under the  assumption that the shape of such a rotating fluid sphere is spheroidal. By introducing the spheroidal coordinates $(x,\eta)$, defined as $\kappa ^2r^2=\left(x^2+c^2\right)\left(1-\eta ^2\right)$, and $\kappa z=x\eta$, where $\kappa $ and $c$ are constants, one can  express the gravitational potential by using spheroidal wave functions.

The effect of the rotation of the dark matter halo in the framework of the BEC dark matter model has also been previously investigated. Two classes of models for rotating haloes were investigated in \cite{inv15}, in order to analyze their stability with respect to vortex formation. In the first model
haloes were modeled as homogeneous Maclaurin spheroids, while in the second one  an $n = 1$ polytropic Riemann-S ellipsoid was considered. Generally, it  was shown that BEC haloes in the polytropic Thomas-Fermi regime typically form vortices.
The dynamics of the rotating Bose Condensate galactic dark matter halos, made of an ultralight spinless boson gas was investigated in \cite{inv16}. The basic approach in this study was to obtain the numerical solution of the system of the coupled Gross-Pitaevskii and Poisson equations, $i\hbar \partial \Psi/\partial t=-\left(\hbar ^2/2m\right)\Delta \Psi +V\Psi+\left(2\pi\hbar ^2a/m^2\right)|\Psi|^2\Psi$ and $\Delta V=4\pi Gm |\Psi |^2$, respectively, in Cartesian coordinates. It was found that ultralight spinless boson dark matter candidates can describe well the galactic astrophysical properties  at local scales with the addition of angular momentum to halos.

In our study we have considered the case in which the halo has an overall rigid body rotation, and we have studied the astrophysically relevant properties that such a rotating halo may have. As a first (and basic) result we have obtained the rotational corrections to the halo density profile, due to rotation. In our investigations we have followed the approach initiated in \cite{chandrasekhar1933equilibrium}, and we have considered the rotation problem in spherical coordinates. While \cite{chandrasekhar1933equilibrium} considers polytropic systems with arbitrary $n$, in the present paper we have systematically investigated the $n=1$ configurations.  In this case the general expression of the density  involves an infinite summation over a set of radial Bessel functions, and angular Legendre polynomials. We have restricted  our  investigations of the astrophysical properties of the BEC halos  to the slow rotation case, when the deformation of the halo is small, and the rotation parameter $\Omega ^2$ satisfies the condition $\Omega ^2<<1$. In this case explicit, and simple expressions of the density of the rotating halo can be obtained. The knowledge of the density distribution is the first step in the investigation of the physical properties of BEC dark matter halos. The mass distribution inside the halo can then be obtained, and the knowledge of the mass profile leads to the expression of the tangential velocity of test particles, following circular orbits around the galactic center. In our approach we have obtained the rotational velocity in the first approximation of $\Omega ^2$, an approximation that may allow the investigation of the effects of the slow rotation on the halo.

The tangential velocity, as well as the density profile, essentially depend on three physical parameters: the central density of the dark matter, the radius of the static density profile, and the angular velocity of the galaxy, respectively. In order to obtain these parameter, we have compared our theoretical results with a (small) set of observational data, and we have fitted our BEC rotating model with 13 observed rotation curves. The sample consisted of 12 dwarf galaxies, and the Milky Way galaxy. The fittings of the 12 dwarf galaxies provided a range of central densities of $\rho _c=\left(0.7-6\right)\times 10^{-24}$ g/cm$^3$, indicating a relatively inhomogeneous distribution of the central densities. The angular velocities also present a relatively large spread, ranging from $0.1\times 10^{-16}$ s$^{-1}$ to $5\times 10^{-16}$ s$^{-1}$, presenting an order of magnitude variation. The predicted masses of the halos also did present a large variation, from $0.17\times 10^{10}$ $M_{\odot}$ to $4.12\times 10^{10}$ $M_{\odot}$, implying a difference by a factor of about 24 between the smallest and the highest galactic mass.

One of the important parameters in the theoretical models of BEC dark matter is the radius $R$ of the static (nonrotating) halo. The value of $R$ is determined by the scattering length $a$ and the mass $m$ of the dark matter particle. As such, $R$ must be a universal constant, and its constancy as proven by the observations, could represent a strong argument in the favor of the condensate dark matter models. In the considered sample of 12 dwarf galaxies $R$ did vary between 2.9 and 8.2 kpc, respectively, by a factor of around 3. However, for this sample of considered galaxies, a {\it fixed} value of $R\approx 6.5$ kpc can give a good fit to all considered observational data. Moreover, such a value of $R$ can give a good fit even to the Milky Way rotation curve, which extends up to 30 kpc, a result which, taking into account the numerous uncertainties in the data and in the (baryonic) fitting model, suggests that the assumption of an  {\it universal value} of $R$ cannot be ruled out by present day observations.

  Our theoretical results have also shown that introducing an angular velocity gives a smaller central density $\rho_c$, and a larger radius at the boundary of the halo. Also, slow rotation slightly increases the total mass of the angular halo. On the other hand, a good knowledge of the baryonic mass distribution in the galaxies, and of the total central mass density can give, via the fitting of the galactic rotation curves, a good indication on whether the halo is rotating, or not, and the value of its angular velocity.

 In the present analysis we did not include the contribution to the gravitational potential of the baryonic galactic matter when we derived the mass distribution of the dark matter halo. This require to modify the Poisson equation to $\Delta V=4\pi G\left(\rho +\rho _b\right)$, where $\rho _b$ is the density of the baryonic matter, and to systematically include the effect of the baryonic matter in the model. In particular, the Helmholtz type equation describing the density distribution of the dark matter would also depend on the density of the baryonic matter. For more luminous galaxies, it may be necessary to take this effect into account.

 The analysis of the galactic rotation curves alone cannot determine the basic physical properties of the condensate dark matter. Alternative physical effects must be taken into account to fully determine the properties of the dark matter particle. One is such physical effects would be the study of the gravitational lensing by BEC halos. It was already shown in \cite{BoHa07} that the BEC dark matter gives a very different prediction for gravitational lensing as compared to other models of dark matter.  One might also include the effects of the vortices in the theoretical analysis, and such an inclusion might explain the wiggles in the rotation curves \cite{inv4}.

 The observation of the possible rotation of the galactic BEC dark matter halos would lead to a deeper understanding of the physics of these complex systems, as well as to some constraints on the nature and physical characteristics of the dark matter particles. In the present investigations we have developed some basic theoretical tools that could help in discriminating between the standard dark matter and condensate dark matter models.

 \section*{Acknowledgments}

Shi-Dong Liang acknowledges the support of the Natural Science Foundation of Guangdong Province
(No. 2016A030313313) and the Project (XJZDL12) of Foreign Expert in Sun Yat-sen University.

\appendix

\section{Derivative of the density profile with respect to $kr$}\label{app}

The derivative with respect to the radial variable $kr$ of the density profile of the Bose-Einstein Condensate dark matter halo, given by the general solution of the Helmholtz equation, can be obtained as follows
\begin{widetext}
\bea
\frac{\mathrm{d}\rho}{\mathrm{d}(kr)}&=&\frac {\mathrm{d}}{\mathrm{d}(kr)}\sum^{\infty}_{l=0}A_{2l}(-kr)^{2l}\left(\frac1{kr}\frac d{d(kr)}\right)^{2l}\frac{\sin kr}{kr}P_{2l}(\cos\theta) \nonumber\\ &&=\frac{\mathrm{d}}{\mathrm{d}(kr)}\sum^{\infty}_{l=0}A_{2l}\sqrt{\frac{\pi}{2kr}}J_{2l+\frac12}(kr)P_{2l}(\cos\theta)\nonumber\\
&& =\sum^{\infty}_{l=0}A_{2l}\Bigg\{-\frac12\sqrt{\frac{\pi}{2(kr)^3}}J_{2l+\frac12}(kr)
 +\sqrt{\frac{\pi}{2kr}}\Bigg[J_{2l-\frac12}(kr)-\frac{2l+1/2}{kr}J_{2l+\frac12}(kr)\Bigg]\Bigg\}P_{2l}(\cos\theta)\nonumber\\
&&=\sum^{\infty}_{l=0}A_{2l}\sqrt{\frac{\pi}{2kr}}\Bigg[\Bigg(-\frac1{2kr}-\frac{2l+1/2}{kr}\Bigg)J_{2l+\frac12}(kr)+J_{2l-\frac12}(kr)\Bigg]P_{2l}(\cos\theta)\nonumber\\
&&=\sum^{\infty}_{l=0}A_{2l}\sqrt{\frac{\pi}{2kr}}\Bigg[-\frac{2l+1}{kr}J_{2l+\frac12}(kr)+J_{2l-\frac12}(kr)\Bigg]P_{2l}(\cos\theta)\nonumber\\
&&=\sum^{\infty}_{l=0}A_{2l}\Bigg[-\frac{2l+1}{kr}j_{2l}(kr)+j_{2l-1}(kr)\Bigg]P_{2l}(\cos\theta).
\eea
\end{widetext}

\end{document}